\documentclass[aps,prl,preprint,groupedaddress]{revtex4-1}

\usepackage{color}
\usepackage{graphicx}
\usepackage{amsfonts}
\usepackage{amsmath}
\usepackage{array}
\usepackage{multirow}
\usepackage{xr}
\usepackage{soul}
\newcolumntype{P}[1]{>{\centering\arraybackslash}p{#1}}
\newcolumntype{L}[1]{>{\raggedright\arraybackslash}m{#1}}

\newcommand{\ket}[1]{\left|#1\right>}
\newcommand{\bra}[1]{\left<#1\right|}

\begin{document}

\title{Comparing and combining measurement-based and driven-dissipative entanglement stabilization}

\author{Y. Liu$^{\ast}$, S. Shankar, N. Ofek, M. Hatridge, A. Narla, K. M. Sliwa, L. Frunzio, R. J. Schoelkopf,  M. H. Devoret$^{\ast}$}
\affiliation{Departments of Applied Physics and Physics, Yale University, New Haven, Connecticut, USA\\
\normalsize{$^\ast$To whom correspondence should be addressed: E-mail:  yehan.liu@yale.edu, michel.devoret@yale.edu}
}

\date{\today}

\begin{abstract}
We demonstrate and contrast two approaches to the stabilization of qubit entanglement by feedback. Our demonstration is built on a feedback platform consisting of two superconducting qubits coupled to a cavity which are measured by a nearly-quantum-limited measurement chain and controlled by high-speed classical logic circuits. This platform is used to stabilize entanglement by two nominally distinct schemes: a ``passive'' reservoir engineering method and an ``active'' correction based on conditional parity measurements. In view of the instrumental roles that these two feedback paradigms play in quantum error-correction and quantum control, we directly compare them on the same experimental setup. Further, we show that a second layer of feedback can be added to each of these schemes, which heralds the presence of a high-fidelity entangled state in real-time. This ``nested'' feedback brings about a marked entanglement fidelity improvement without sacrificing success probability. 
\end{abstract}

\maketitle


\section{Introduction}

The ability to perform quantum error correction (QEC) by feedback is a crucial step towards fault-tolerant quantum computation \cite{Nielsen2004, Terhal2015quantum}. An open challenge, that has drawn considerable interest recently\cite{Kerckhoff2010,Fowler2012,Fujii2014}, is to find the best strategy for this task. Two nominally distinct feedback strategies for QEC are the measurement-based and driven-dissipative approaches. The former has been more well-understood\cite{Wiseman2009}, owing to an existing foundation in classical control and feedback in engineering. In the measurement-based (MB) approach, a classical controller performs projective measurements of a set of multi-qubit stabilizer operators that encode the logical qubit\cite{Nielsen2004,Fowler2012} in order to track errors and/or perform any necessary correction. Thus for good performance, this approach requires both high-fidelity projective measurements and low-latency control electronics to process the measurement result within the relevant coherence times of the quantum system. The elements required for this MB strategy have been demonstrated for small quantum systems on various physical platforms such as Rydberg atoms\cite{Haroche2011Nature}, trapped ions\cite{Chiaverini2004, Nigg2014quantum}, photons\cite{Yao2012}, spin\cite{Waldherr2014} and superconducting qubits\cite{Riste2012b,Vijay2012,Campagne-Ibarcq2013,Riste2013Nature,Steffen2013Nat,Barends2014Nature, Chow2014Nature, Riste2014NatComm, Kelly2015Nature, Corcoles2015}. However, a steady-state multi-qubit QEC capability has yet to be achieved and one of the key questions for this development is whether the MB strategy is scalable to larger systems or whether an alternative approach is more optimal.

One such alternative, driven-dissipative (DD) schemes\cite{Poyatos1996}, also called reservoir/bath engineering or coherent feedback (as discussed below), utilizes coupling between the quantum system of interest and a dissipative environment to transfer the entropy caused by decoherence-induced errors out of the quantum system. They have been demonstrated on a variety of physical systems including atomic ensembles\cite{Krauter2011}, trapped ions\cite{lin2013dissipative}, mechanical resonators\cite{Kerckhoff2013,kienzler2015} and superconducting qubits\cite{Murch2012, Geerlings2013, Leghtas2013, Shankar2013,Leghtas2015, holland2015}.  Moreover, experiments with trapped ions\cite{Schindler2011} and superconducting qubits\cite{Reed2012} have demonstrated some of the basic elements of autonomous QEC.  These schemes do not require high-fidelity projective measurements, external control and its associated latency. They can also be described as autonomous or coherent feedback\cite{Wiseman1994}, where the reservoir coupled to the target quantum system can be considered as an effective ``quantum controller'' that reacts with quantum degrees of freedom\cite{Nurdin2009}. Adjusting the feedback by changing the ``quantum controller'', however, can be more challenging than re-programming a classical controller, built with conventional electronics. Thus, a further question is whether one can combine this DD approach and the conventional MB approach with minimal negative consequences from their respective drawbacks.

Here we report an experiment in which we built a feedback platform utilizing a nearly quantum-limited measurement chain and a customized field-programmable gate array (FPGA) system to perform MB and DD schemes within the same setup. The task of this platform was to stabilize an entangled Bell state of two superconducting transmon qubits\cite{Schreier2008}. This particular task of stabilizing a single state is a proxy for more general QEC experiments where a manifold of states is protected. We realize, for the first time, an MB \textit{stabilization} of a Bell state by repeated active correction through conditional parity measurements\cite{Lalumiere2010, Tornberg2010,Riste2013Nature}. We compare this scheme to a DD entanglement stabilization scheme\cite{Shankar2013} in which the conditional parity switch is autonomous. By performing both schemes on the same hardware setup and circuit QED (cQED) system\cite{Blais2004}, we shed light on their close connection and compare them on a level playing field. 

Previous theoretical works have compared DD (under the name of ``coherent feedback'') and MB for linear quantum control problems\cite{Yamamoto2014coherent}, such as for minimizing the time required for qubit state purification\cite{Jacobs2014coherent} or for cooling a quantum oscillator\cite{Hamerly2012advantages}. These comparisons  showed coherent feedback to be significantly superior. In our particular setup, we find that distinguishing the superior approach among DD and MB is a more subtle task. The subtlety is two-fold. First, the performance difference depends on which process can be better optimized: the design of the cQED Hamiltonian or the efficiency of quantum measurement and classical control. In the current experiment, we show that DD has better steady-state performance as the cQED Hamiltonian parameters are engineered such that DD has a shorter feedback latency. But DD's advantage over MB is not immutable. As certain experimental parameters are improved, such as coherence times and measurement efficiency, MB's performance can catch up with DD. 

Secondly, in the current situation in which neither the cQED Hamiltonian parameters nor the measurement and control parameters are ideal, we can obtain a boosted performance by combining DD and MB to get the best of both worlds. We explored this by devising a heralding method to improve the performance of both stabilization approaches. This protocol exploits the high-fidelity measurement capability and the programmability of the feedback platform. The protocol is termed ``nested feedback'' since it has an inner feedback loop based on either the DD or MB scheme, and an outer loop that heralds the presence of a high-fidelity entangled state in real-time. Previously, heralding schemes have been demonstrated for state preparation to combat photon loss or decoherence\cite{Moehring2007,Wagenknecht2010,Hofmann2012heralded,Johnson2012, Riste2012, Riste2012b, Riste2013Nature,Bernien2013heralded}. Extending such heralding capability to state stabilization will be a valuable addition to the QEC toolbox. Furthermore, the ability to herald in real time as opposed to post-selection is important for on-demand and deterministic quantum information processing since only successful events lead to subsequent processing. Real-time heralding for entanglement stabilization is particularly challenging for superconducting qubits due to their shorter coherence times compared to other systems. In this article, we implement this real-time heralding capability on a time scale faster than the few microsecond coherence time of our qubit-cavity system. By extending the feedback platform developed primary for the MB approach to the DD approach, our results bring to light a new application of MB. Adding a level of MB feedback can significantly improve performance beyond what a single layer of feedback, whether DD or MB, can achieve.

\section{Experiment setup}

\begin{figure}
\centering
\includegraphics{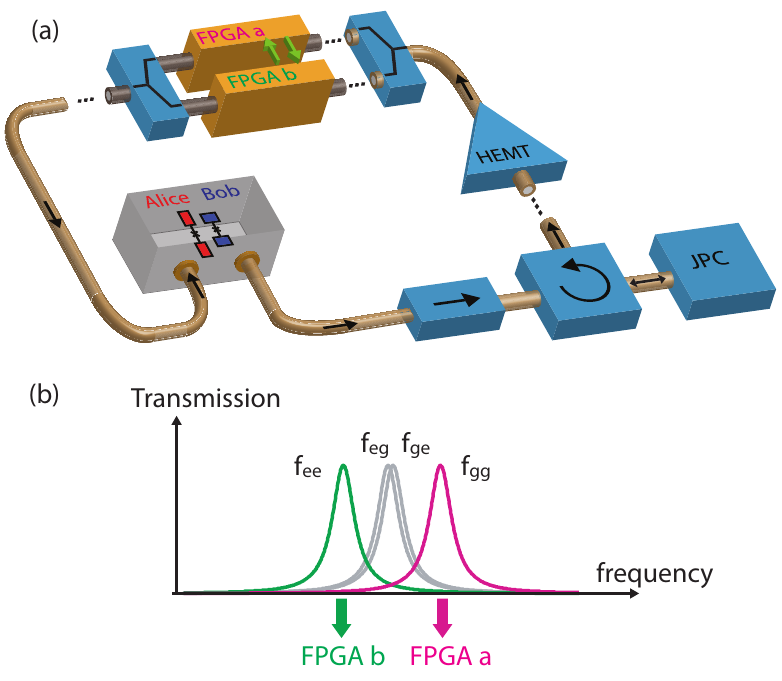}
\caption{\label{fig:schematic}(a) Schematic of the experimental set-up. Two independently addressable transmon qubits, Alice and Bob, are dispersively coupled to a three-dimensional microwave cavity. The cavity output is directed to a nearly quantum-limited measurement chain consisting of a Josephson amplifier (JPC) followed by a semiconductor amplifier (HEMT). A pair of custom Field-Programmable Gate Array boards (FPGA a, b) monitor the amplified output and generate real-time modulated microwave drives to control the cavity-qubit system. (b) Transmission spectra of the cavity. Cavity outputs at $f_{gg}$ and $f_{ee}$ are fed to FPGA a and b, respectively.}
\end{figure}

The simplified schematic of our experimental setup is shown in Fig.~\ref{fig:schematic}a. Two transmon qubits\cite{Schreier2008}, Alice and Bob, are dispersively coupled to a three-dimensional aluminum cavity \cite{Paik2011}, with linewidth $\kappa/2\pi = 2$~MHz (see Supp. Mat. Sec.~I, for other parameters).  The qubit-cavity dispersive shifts are nearly equal and in the strong dispersive regime ($\chi_{\text{Alice}}/2\pi = 5$~MHz, $\chi_{\text{Bob}}/2\pi = 4.5$~MHz) with photon-number resolved qubit transition frequencies\cite{Schuster2007}. The cavity output is amplified by a Josephson Parametric Converter (JPC) operated as a nearly quantum-limited phase-preserving amplifier \cite{Bergeal2010a} enabling rapid, single-shot readout\cite{Hatridge2013} and thus real-time feedback. The key component of the experiment is a controller realized with two FPGA boards \footnote{X6-1000M from Innovative Integration} that both measure and actively control the cavity-qubit system. 

An essential operation for our experiment is a two-qubit joint quasi-parity measurement using the common readout cavity \cite{Tornberg2010, Lalumiere2010, Riste2013Nature}. As shown in Fig.~\ref{fig:schematic}b, the cavity is driven at $f_{gg}$ (both qubits in ground state) and at $f_{ee}$ (both in the excited state) at the same time. The output at $f_{gg}$ and $f_{ee}$ together distinguishes the even parity manifold $\{\ket{gg},\ket{ee}\}$ from the odd parity manifold $\{\ket{ge},\ket{eg}\}$. When the two cavity output responses \textit{both} have an amplitude below a certain threshold, the qubits are declared to be in odd parity; when either one has amplitude above the threshold, the qubits are declared to be in even parity. We note that, unlike a true parity measurement, this readout actually distinguishes the two even parity states $\ket{gg}$ and $\ket{ee}$, hence we refer to it as a ``quasi'' parity measurement.  However, the feedback schemes described below apply the same operation on both even states, and thus we need only record the parity of the measured state. The choice of driving at the ``even'' cavity resonances rather than between the ``odd'' resonances ($f_{eg}$ and $f_{ge}$) mitigates the effect of the $\chi$ mismatch, reducing associated measurement-induced dephasing of the odd manifold\cite{Tornberg2010}. The controller FPGA a (b) modulates the $f_{gg}$ ($f_{ee}$) drive to the cavity and also demodulates the response. The two FPGAs share their measurements of the cavity response to jointly determine the parity. In addition, FPGA a and b generate the qubit pulses to Alice and Bob, respectively, which are conditioned on the joint state estimation during real-time feedback.

\section{Principle of experiment and results}

\begin{figure}
\centering
\includegraphics{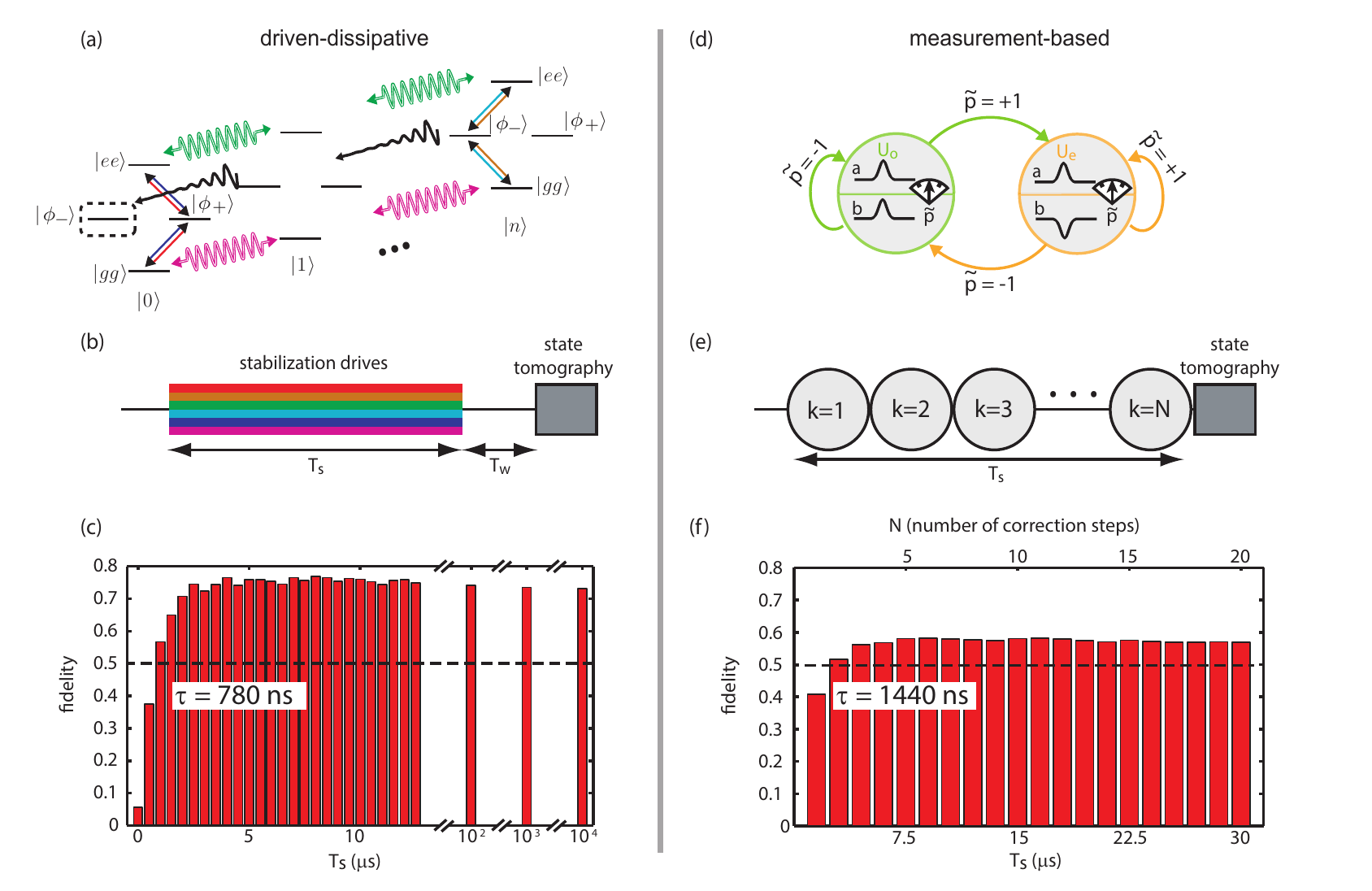}
\caption{\label{fig:comparison}
}
\end{figure}

We first briefly outline the DD stabilization of entanglement, described in detail in Ref.~\onlinecite{Leghtas2013} and \onlinecite{Shankar2013}. This stabilization targets the two-qubit Bell state $\ket{{\phi}_{-}}=\frac{1}{\sqrt{2}}(\ket{ge}-\ket{eg})$.  Figure ~\ref{fig:comparison}a displays the states coupled by the autonomous feedback loop.  Two Rabi drives on Alice and Bob at their zero-photon qubit frequencies ($\omega_{\text{Alice}}^0$ and $\omega_{\text{Bob}}^0$, see Supp. Mat. Sec.~I) couple the wrong Bell state $\ket{{\phi}_{+}}$ to the even states, $\ket{gg}$, $\ket{ee}$, in the energy manifold with zero cavity photons. A second pair of Rabi drives at the $n$-photon qubit frequencies ($\omega_{\text{Alice}}^0-n\bar\chi$ and $\omega_{\text{Bob}}^0-n\bar\chi$, $\bar\chi = \left(\chi_{\text{Alice}}+\chi_{\text{Bob}}\right)/2$), with their relative phase opposite to the first pair, couple $\ket{gg, n}$, $\ket{ee, n}$ to the Bell state $\ket{{\phi}_{-}, n}$. The two cavity drives, at $f_{gg}$ and $f_{ee}$ connect the two manifolds and hence the combined action of the six drives transfers the population from $\ket{gg}$, $\ket{ee}$ and $\ket{{\phi} _{+}}$ to $\ket{{\phi}_{-}, n}$. Finally, cavity photon decay brings $\ket{{\phi}_{-}, n}$ back to $\ket{{\phi}_{-}, 0}$. In effect, the cavity drives separate qubit states based on their parity, allowing one pair of Rabi drives to move the erroneous odd population to the even states while the other pair transfers the even states population to $\ket{{\phi}_{-}}$. 

Counterparts to these elements of the DD feedback loop can be found in the corresponding MB feedback scheme. The action of our MB algorithm is shown as a state machine in Fig.~\ref{fig:comparison}. We describe the quasi-parity measurement $\tilde{P}$ by the projectors $P_{odd} = \ket{ge}\bra{ge} + \ket{eg}\bra{eg}$, $P_{gg} = \ket{gg}\bra{gg}$ and  $P_{ee} = \ket{ee}\bra{ee}$. We assign the outcomes  $\tilde{p} = +1$ to the even projectors, $P_{gg}$ and $P_{ee}$ and $\tilde{p} = -1$ to $P_{odd}$. The MB algorithm is built with a sequence of correction steps, each of which consists of a conditional unitary and a quasi-parity measurement.  The two possible states of the state machine correspond to whether we apply the unitary $U_E$ or $U_O$, followed by the quasi-parity measurement. Specifically, $U_E=R^{\text{a}}_{\text{x}}(\frac{\pi}{2}) \otimes R^{\text{b}}_{-\text{x}}(\frac{\pi}{2})$ where a (b) denotes Alice (Bob), and $U_O = R^{\text{a}}_{\text{x}}(\frac{\pi}{2}) \otimes R^{\text{b}}_{\text{x}}(\frac{\pi}{2})$. In a correction step $k$, the qubits are initially in either $\ket{gg}$, $\ket{ee}$  or in the odd manifold, due to the projective quasi-parity measurement in step $k-1$; the controller then applies $U_E$ ($U_O$) if $\tilde{p}$ in previous step reported $+1$ ($-1$).

The effect of the state machine on the two-qubit states is shown in Tab.~\ref{tabu:mb}, where the action of the controller during one correction step is described in terms of the four basis states, $\ket{{\phi}_{-}}$, $\ket{{\phi}_{+}}$, $\ket{gg}$ and $\ket{ee}$ (the latter two are grouped in the "even" column). The quasi-parity measurement infidelity, labeled by $\epsilon_{E|O}$ ($\epsilon_{O|E}$), gives the error probability of obtaining an even (odd) parity outcome after generating an odd (even) state. Because these measurement infidelities are small, the dominant events are those that occur without measurement errors. At each step, $U_E$ on either $\ket{gg}$ or $\ket{ee}$ followed by the quasi-parity measurement $\tilde{P}$ transfers the states to $\ket{\phi_-}$ with 50\% probability.  Since $\ket{{\phi}_{-}}$ is an eigenstate of $U_O$ and $\tilde{P}$ (modulo a deterministic phase shift that can be undone, see later discussion), these operations leave it unaffected.  On the other hand, $U_O$ and $\tilde{P}$ transform $\ket{\phi_+}$ into $\{\ket{gg},\ket{ee}\}$; more generally, they take population in any other odd state (i.e., a superposition of $\ket{{\phi}_{-}}$ and $\ket{\phi_+}$ ) into $\ket{{\phi}_{-}}$ and the even states.

\begin{table}
\begin{tabular}{L{4cm} | P{1.8cm} | P{1.8cm} | P{1.8cm} | P{1.8cm}|  P{0.9cm} | P{0.9cm} | P{0.9cm} | P{0.9cm}}
	\hline
	\hline
	Previous state & \multicolumn{2}{c|}{$\ket{\phi_-}$} & \multicolumn{2}{c|}{$\ket{\phi_+}$} & \multicolumn{4}{c}{even} \\ \hline
	$\tilde{p}_{k-1}$ & ${+1}$ & ${-1}$ & ${+1}$ & ${-1}$ & \multicolumn{2}{c|}{${+1}$} & \multicolumn{2}{c}{${-1}$} \\ \hline
	Outcome probability & $\epsilon_{E|O}$& $1- \epsilon_{E|O}$ & $\epsilon_{E|O}$ & $1- \epsilon_{E|O}$  & \multicolumn{2}{c|}{$1- \epsilon_{O|E}$}  & \multicolumn{2}{c}{$\epsilon_{O|E}$}  \\ \hline 
	Unitary & $U_E$ & $U_O$ & $U_E$ & $U_O$ & \multicolumn{2}{c|}{$U_E$} & \multicolumn{2}{c}{$U_O$} \\ \hline
	Next state & even & $\ket{\phi_-} $ & $\ket{\phi_+}$ & even & \multicolumn{2}{c|}{even/$\ket{\phi_-}$} & \multicolumn{2}{c}{even/$\ket{\phi_+}$} \\ \hline
	\hline
\end{tabular}
\caption{\label{tabu:mb} Effects of the MB finite state machine of Fig.~2 on two-qubit system in the $k$-th step of feedback, for different starting cases (columns). Row 2 through 4 describe the result of the previous quasi-parity measurement and the corresponding unitary that will be applied in the $k$-th step. The symbols, $\epsilon_{E|O}$ and $\epsilon_{O|E}$, denote parity measurement errors (see text).  The last row describes the possible system states attained by the applied unitary. The two alternative states for a previous ``even'' occur with $50\%$ probability.}
\end{table}

By repeating a sufficient number of these correction steps in sequence, the controller stabilizes the target Bell state irrespective of the initial two-qubit state.  The similarity between this active feedback and DD is that MB also transfers population between different parity states by conditional Rabi drives. However, while the Rabi drives in DD are conditioned autonomously by the photon number in the cavity, the unitary Rabi pulses in MB are conditioned by real-time parity measurement performed by active monitoring of cavity outputs.

The pulse sequences for DD and MB are shown in Fig.~\ref{fig:comparison}b and e. In DD, a set of continuous-wave drives are applied for a fixed time $T_{\text{s}}$ and after some delay $T_{\text{w}}$ to allow remaining cavity photons to decay, a two-qubit state tomography is performed \cite{Filipp2009,Chow2010}. The cavity and Rabi drive amplitudes and phases were tuned for maximum entanglement fidelity, following the procedure described in Ref.~\onlinecite{Shankar2013}. In particular, the optimal cavity drive amplitudes were found to be $\bar n = 4.0$. For MB, the continuous drives are replaced by a pre-defined number of correction steps $N$, resulting in a stabilization duration of $T_s = N T_{step}$ where $T_{step} = 1.5$~$\mu$s. There is no extra delay before tomography since each correction step already contains a delay after the quasi-parity measurement due to feedback decision latency. The strength and duration of the quasi-parity measurement $\tilde{P}$ were optimized as discussed in Supp.~Mat.~Sec.~II. The optimization achieved low parity measurement infidelities $\epsilon_{E|O}$ and $\epsilon_{O|E}$ while keeping the measurement-induced dephasing arising from the $\chi$ mismatch\cite{Tornberg2010,Lalumiere2010} small compared to the natural decoherence in the same duration. We experimentally determined the infidelity of the quasi-parity measurement to be $\epsilon_{E|O}$ and $\epsilon_{O|E}$ of $0.04$ and $0.05$, respectively. The quasi-parity measurement also causes a deterministic qubit rotation about the respective $Z$ axis due to an AC Stark shift~\cite{Tornberg2010}; this rotation was corrected within the unitary gate $U_O$ as discussed in Supp.~Mat.~Sec.~VI. 

Fig.~\ref{fig:comparison}c,f show the fidelity to the target Bell state $\ket{{\phi}_{-}}$ as a function of stabilization time for DD and MB, respectively. The fidelity rises exponentially with a characteristic time constant of $0.78$~$\mu$s ($1.4$~$\mu$s) and a steady-state fidelity of $76\%$ ($57\%$) for DD (MB). Both fidelity values agree with numerical modeling based on master equation simulation, which gives $76\%$ and $58\%$ for DD and MB, respectively (see Supp.~Mat.~Sec.~IV). The exponential dependence is a signature of feedback and arises from the characteristic loop time. The experimentally determined time constants are in reasonable agreement with their simulated values of $1.0$~$\mu$s ($1.4$~$\mu$s) for DD (MB). In MB, this loop time is related to the step length ($1.5$~$\mu$s), which is given by the sum of the quasi-parity measurement duration ($0.66$~$\mu$s), the cable, instrument and FPGA latencies ($0.69$~$\mu$s), and the duration of unitary pulses ($0.15$~$\mu$s). On the other hand for DD, the measured loop time is close to $10$ cavity lifetimes, the expected time as shown in Ref.~\onlinecite{Leghtas2013}.

The superior performance of DD over MB for the steady-state fidelity is due to the difference in correction loop time, which needs to be shorter than the coherence times of the two qubits for high fidelity entanglement. For the current experimental setup, the latency of the controller and quantum efficiency of the measurement chain, which affects the fidelity of the single-shot readout, result in a longer loop time in MB. A source of the longer feedback loop time is the quasi-parity measurement duration. This measurement duration, which was optimized as discussed in Supp.~Mat.~Sec.~II, is limited by dephasing induced by the mismatch in $\chi$ ($\sim 10\%$) and the measurement efficiency of the output chain ($\sim 30\%$), which can both be improved in future experiments. Our simulations (see Supp.~Mat.~Sec.~IV) suggest that with current state-of-the-art measurement efficiency value and optimization of the FPGA/cable latency, the MB steady state fidelity can be improved to $66\%$. The limited measurement efficiency does not affect the performance of DD because the parity measurement and correction take place autonomously within the qubit-cavity system, indicative of its robustness against this hardware limitation. On the other hand, both DD and MB schemes benefit from longer intrinsic coherence times and reduction of the $\chi$ mismatch. For example, simulations show that if the coherence times are improved to one hundred microseconds (achieved in other state-of-the-art cQED setups), both DD and MB fidelities can increase to above 85\%. For the rest of the article, however, we consider boosting the fidelity in a different manner, without making any physical changes to the qubit-cavity system.

The DD and MB schemes described so far are synchronous in the sense that the stabilization always ends after a pre-determined duration and the tomography follows. Decoherence causes the qubits to have a finite probability of jumping out of the target state immediately before the stabilization terminates. Hence, this ``fixed time'' protocol does not always output the target state with maximum fidelity. An optimum protocol would rather utilize all available information. For both DD and MB, we can measure the cavity output at the end of the stabilization period. The outcomes of these measurements, $I_{gg}$ and $I_{ee}$, give real-time information on the state of the two qubits, and thus can herald a successful stabilization sequence.

\begin{figure}
\centering
\includegraphics{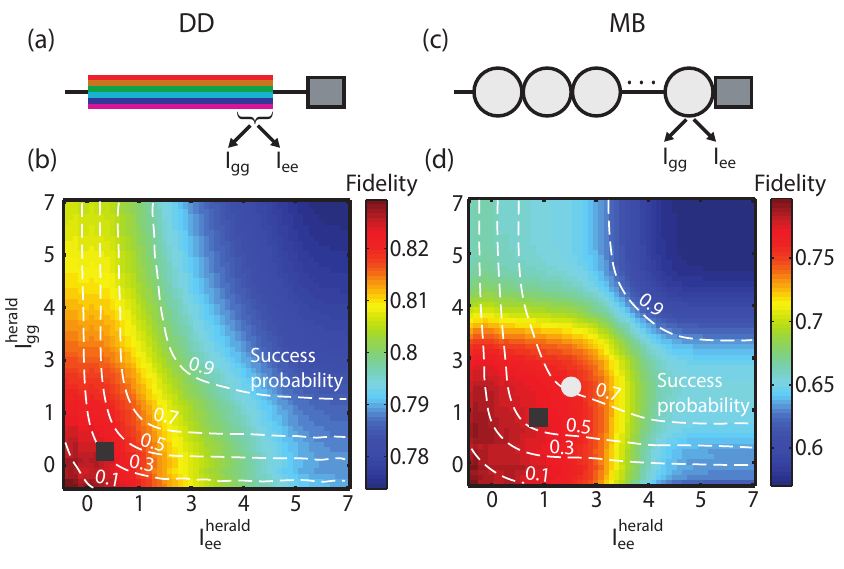}
\caption{\label{fig:thresh_sweep} Trade-off between success probability and fidelity for both DD and MB schemes. (a) Pulse sequence for heralding DD stabilization by post-selection. At the end of the stabilization period, the cavity outputs, $I_{gg}$ and $I_{ee}$ are measured at their respective frequencies. (b) Color plot of fidelity to the target state for DD as a function of thresholds chosen for $I_{gg}$ and $I_{ee}$ (see supplementary material for details of the thresholds). Also plotted as white dashes are contour lines of the success probability associated with each choice. Solid black square indicates the thresholds chosen for the condition $C$ in the nested feedback protocol described in Fig.~4. (c) and (d) same as above for MB, including the corresponding thresholds for $C$. Gray circle indicates the thresholds chosen for the quasi-parity measurement $\tilde{p}$ used to condition MB correction steps.}
\end{figure}

In Fig.~\ref{fig:thresh_sweep}, we describe how monitoring the cavity outputs improves target state fidelity. We introduce two thresholds $\{I_{gg}^{herald},I_{ee}^{herald}\}$ (see supplementary material for details) to post-select the measurement outcomes of $I_{gg}$ and $I_{ee}$ respectively, and identify successful stabilization runs \cite{Riste2013Nature, Shankar2013}. The results of varying $\{I_{gg}^{herald},I_{ee}^{herald}\}$ are shown in Fig.~\ref{fig:thresh_sweep}b,d for DD and MB, respectively. The color plots show fidelity improving as the thresholds become more stringent. The success probability defined as the percentage of stabilization runs kept for tomography given a set of thresholds, is also plotted as contours for both DD and MB. There is a clear trade-off between success probability and fidelity. To reach the maximum fidelity in DD of 82\%, at least 75\% of experiment runs need to be discarded. The trade-off is less severe in MB, where only 50\% of runs need to be discarded to reach the maximum fidelity of 75\%. However we aim to eliminate this trade-off all together, i.e., to improve the fidelity while maintaining a high success probability.

\begin{figure}
\centering
\includegraphics{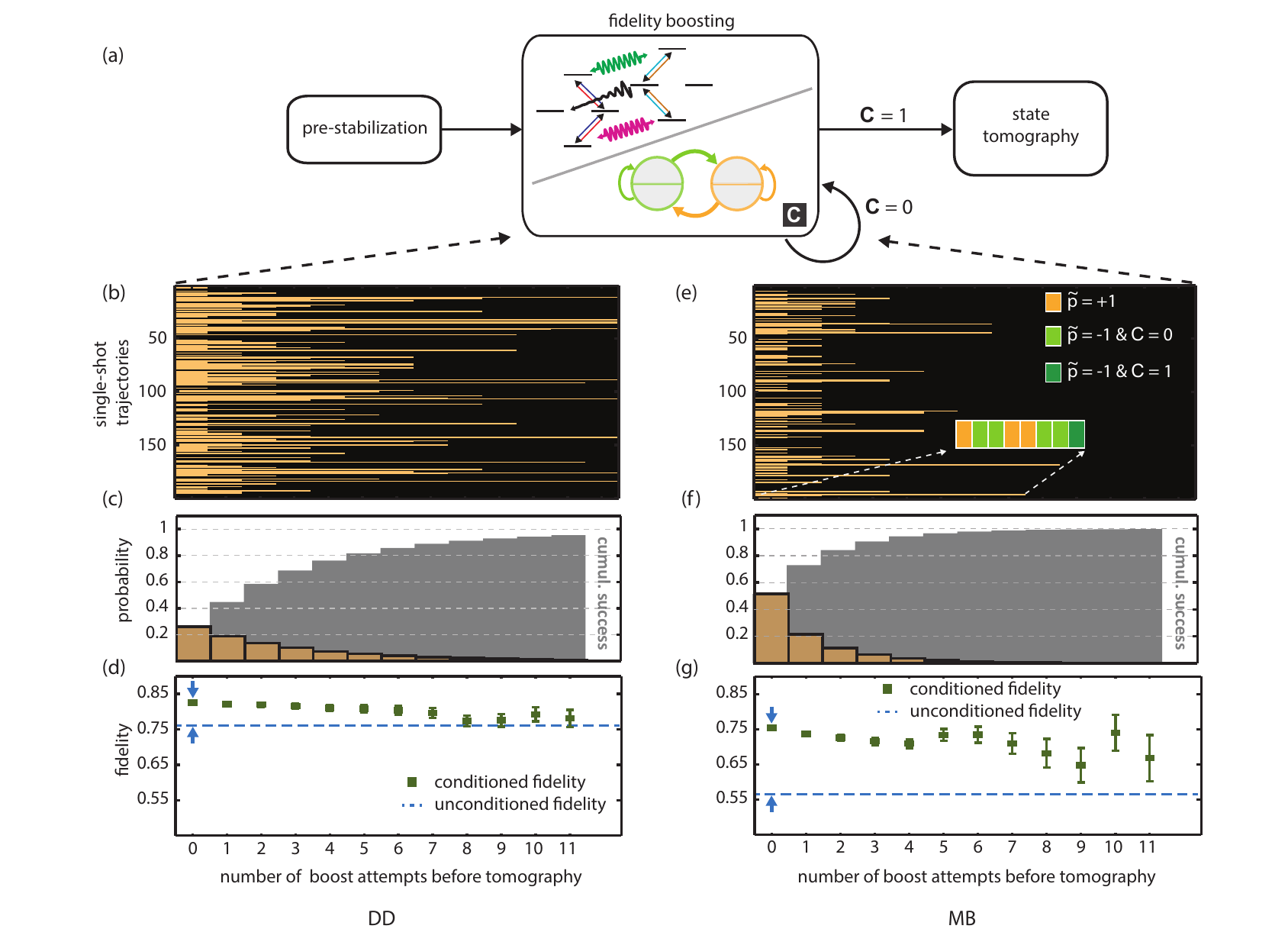}
\caption{\label{fig:real_time}  Nested feedback protocol for boosting fidelity and heralding successful stabilization run in real-time. (a) State machine representation. The control variable $C$ (see main text and Fig.~3) determines the repetition of a boost cycle or the heralding of a successful run. The maximum number of boost attempts allowed is set to 11 in the experiment. (b), (e) 200 single-shot sequence trajectories of nested feedback for DD and MB, respectively. The trajectories are colored yellow during boost attempts and black after $C$ is satisfied. Trajectories that are entirely black satisfied the success criterion without the need for boost.  The inset in (e) shows an example of an MB trajectory consisting of both $\tilde{p}$ and $C$ outcomes. (c),(f) Cumulative success probability (gray shade) of having completed at most a given number of boost attempts before tomography for DD and MB, respectively. Yellow bars indicate differential success probability. (d),(g) Fidelity to target Bell state for DD and MB, respectively. Green squares show the corresponding fidelity as a function of the number of boost attempts. The blue dashed line denotes fidelity without any boosting (i.e.\ unconditioned by $C$). Blue arrows denote the improvement in fidelity due to nested feedback. While the absolute fidelity of MB is worse than DD due to latency, the fidelity boost is higher.}
\end{figure}

This goal is achieved by introducing a nested feedback protocol (NFP), in which the stabilization feedback loop enters into a higher layer of feedback for ``fidelity boosting'' instead of proceeding to state tomography directly. In contrast to the ``fixed time'' protocol, NFP conditions the termination of stabilization on the quality of the entanglement, i.e., it heralds a successful stabilization run in real-time, as illustrated by the state machine diagram in Fig.~\ref{fig:real_time}a. The control variable $C$ is given by $C = (I_{gg} < I_{gg}^{herald})\ \textbf{AND}\ (I_{ee} < I_{ee}^{herald})$, where $\{I_{gg}^{herald},I_{ee}^{herald}\}$ are determined by the same post-selection experiment discussed previously to optimize the fidelity (black square in Fig.~3b and d). If the controller determines that the entanglement quality is not sufficient ($C = 0$), a boost phase is attempted which comprises exactly one correction step for MB or a stabilization period of similar duration for DD ($1.4$~$\mu$s). During the boost phase, the cavity outputs are integrated to give $\{I_{gg},I_{ee}\}$ which enables the next real-time assessment of $C$. In DD, the parity measurement and first layer of feedback is accomplished autonomously, therefore the FPGA only needs to check $C$. However in the MB scheme both layers of feedback are performed solely by the FPGA. It therefore checks if $C=1$ to herald that the entanglement meets the desired quality. If not, it uses the quasi-parity thresholds (grey circles in Fig.~3d) to decide whether the qubits are in even or odd state in order to continue stabilization. This asynchronous pulse sequencing and conditioning by multiple thresholds exploit the programmable nature of the FPGA-based platform.

The asynchronous behavior of NFP is displayed in Fig.~\ref{fig:real_time}b(e) for DD (MB), which demonstrates 200 single-shot runs. The DD (MB) fidelity boosting sequence continues until either success or a maximum limit on boost attempts (set to 11 in the experiment) is reached. For the MB protocol, the trajectory of the qubits' parity can be tracked by the conditioning outcomes of the inner-loop control variable $\tilde{p}$ and the outer-loop control variable $C$, which are independent.  Through repeated boost attempts until success, NFP significantly improves the overall success probability. Within 11 attempts, 95\% (99.8\%) of DD (MB) runs satisfy the success condition compared to just 25\% (50\%) with simple post-selection. This is assessed by the cumulative probability, the integral of the probability of having completed a certain number of boost attempts before tomography, as plotted in Fig.~\ref{fig:real_time}c,f. Since MB requires a less stringent threshold than DD to gain fidelity improvement, the MB success probability converges to unity much faster than that of DD. Finally, we show that the high success probability does not come at the cost of reduced fidelity. The fidelity to $\ket{{\phi}_{-}}$ for DD improves from an unconditioned value of 76\% to 82\% (averaged over all successful attempts). For MB, the improvement is more pronounced: fidelity rises from an unconditioned value of 57\% to 74\%. Thus for both DD and MB, NFP attain close to the fidelity achieved via stringent post-selection. These results for NFP also agree well with a numerical simulation (see Supp.~Mat.~Sec.~V).

One will note, however, a continuous downward trend of the fidelity in both DD and MB schemes as the number of attempts increases. This is due to the non-negligible population in the $\ket{f}$ states of the two qubits in the experiment, which escape correction by the stabilization feedback loops. After each further boost attempt of stabilization, the probability of the population escaping outside the correction space thus increases, diminishing the fidelity (see Supp.~Mat.~Sec.~VII). Also note that the error bars on the fidelity of MB are bigger than those in DD for large attempt numbers simply because the probability of needing many attempts is lower in MB than in DD. 

While real-time heralding by NFP removes the trade-off between fidelity and success probability, it does so by introducing a different trade-off -- high fidelity and success probability are achieved but the protocol length now varies from run to run. If NFP is a module within a larger quantum information processing (QIP) algorithm, then this asynchronous nature must be accommodated by the controller. For our FPGA-based control, NFP is easily accommodated because it is a natural extension to ``fixed-time'' or synchronous operation. In ``fixed time'' operation, the controller conditions its state by the protocol length which is pre-determined and stored in an internal counter by the experimenter. On the other hand in NFP, the controller conditions its state on a pre-determined logical function of its real-time inputs.

\section{Conclusion}
In conclusion, we have implemented a new MB stabilization of an entangled state of two qubits, which parallels a previous DD stabilization scheme. Instead of coherent feedback by reservoir engineering, MB relies on actively controlled feedback by classical high-speed electronics external to the quantum system. When comparing both schemes in the ``fixed-time'' protocol, we observe that DD gives a higher fidelity to the target state due to lower feedback latency. Furthermore, we have improved the fidelity of both schemes by a nested feedback protocol which heralds stabilization runs with high-quality entanglement in real time. The real time heralding brings about the fidelity improvement without a common trade-off in QIP: it does not sacrifice the experiment success probability. It eliminates this trade-off by allowing asynchronicity in the experiment.

Our experiment shows some of the key advantages of MB platforms that have not been previously explored. Typically, the performance of MB feedback has not been at par with methods based on post-selection, due to the latency of the controller. However it is widely recognized that the trade-off of success probability for fidelity in the case of post-selection is untenable for large scale systems. Therefore, existing digital feedback\cite{Campagne-Ibarcq2013, Riste2012b, Riste2013Nature,Steffen2013Nat} have focused on achieving nearly perfect success probability. Here, we are exploring another direction of feedback which achieves high fidelity with high success probability. Our nested feedback strategy maximizes the use of the information coming out of the qubit-cavity system in order to make the correction process as efficient as possible. We find that our feedback platform, comprised of a nearly-quantum-limited measurement chain and a real-time classical controller, provides the necessary tool-set to implement such a strategy. We show that this technology can be extended to improve the performance of DD approaches as well as single-layer MB approaches themselves. This strategy could be carried out further in the future. For example, the FPGA state estimator could perform a more sophisticated quantum filter of the microwave output of the DD stabilization to herald successful events with better accuracy, significantly improving the success probability convergence rate.

Similar ideas can be applied in the future towards other forms of stabilization, such as for stabilizing Schr{\"o}dinger cat states of a cavity mode\cite{Mazyar2014}, a proposed logical qubit. Initial experiments on such logical qubits with high fidelity-measurement\cite{Sun2014} or dissipation engineering\cite{Leghtas2015} have been performed and could now be combined. Likewise, future logical qubits based on the surface code\cite{Fowler2012} could also be stabilized by either active stabilizer measurements\cite{Barends2014Nature, Chow2014Nature, Riste2014NatComm} or as recently proposed by dissipation engineering\cite{Kapit2015PRA,Fujii2014}. Our experiment demonstrates that measurement-based and driven-dissipative approaches, far from being antagonistic, can be merged to perform better than either approach on its own.

\section{Acknowledgements}
We thank Zaki Leghtas, Mazyar Mirrahimi, Matti Silveri and Shantanu Mundhada for helpful discussions. This research was supported by the U.S. Army Research Office (W911NF-14-1-0011 and W911NF-14-1-0563). Facilities use was supported by the Yale Institute for Nanoscience and Quantum Engineering and NSF MRSEC DMR 1119826.  All statements of fact, opinion or conclusions contained herein are those of the authors and should not be construed as representing the official views or policies of the U.S. Government.

\clearpage

\clearpage

\clearpage
FIG.~2. Comparison between the driven-dissipative (DD) and the measurement-based (MB) entanglement stabilization, shown in the left (DD) and right (MB) panels.  (a) Diagram of qubit/cavity state evolution in the DD feedback loop. Two-qubit state manifolds are laddered for different photon numbers in the cavity (labeled by $\ket{n}$). The green and pink sinusoidal double arrows represent cavity drives, while the straight double red/blue and cyan/yellow arrows are Rabi drives on the qubits. These six drives and the cavity dissipation (black decaying arrow) couple the different states of the system such that the target state $\ket{{\phi}_{-}}$ is stabilized. (b) Functional pulse sequence for DD. The duration needed to empty the cavity of residual photons (see text) before tomography is indicated by $T_{\text{w}}$. (c) Fidelity to the target as a function of stabilization duration ($T_{\text{s}}$). Dashed line at $0.5$ denotes the threshold for entanglement. The time given in the white box, $\tau$, is the characteristic time constant of the exponential rise of fidelity. (d) State machine representation of the MB feedback loop. The quasi-parity measurement reports $\ket{gg}$, $\ket{ee}$ as $\tilde{p}=+1$ (even) and $\ket{{\phi}_{-}}$, $\ket{{\phi}_{+}}$ as $\tilde{p}=-1$ (odd). For odd (even) parity, two $\frac{\pi}{2}$ pulses, with identical (opposite) phases, are applied to Alice and Bob, respectively. (e) Sequence of correction steps conditioned by the quasi-parity measurement and leading into tomography. Counter $k$ limits the number of steps to $N$. (f) Fidelity to the target Bell state as a function of stabilization duration ($T_{S}$) or number of correction steps ($N$).
\clearpage

\clearpage

\clearpage

\pagebreak
\widetext
\begin{center}
\textbf{\large Supplementary materials}
\end{center}
\setcounter{equation}{0}
\setcounter{figure}{0}
\setcounter{table}{0}
\setcounter{page}{1}
\setcounter{section}{0}

\makeatletter
\renewcommand{\thesection}{\Roman{section}}
\renewcommand{\theequation}{S.\arabic{equation}}
\renewcommand{\thefigure}{S.\arabic{figure}}
\renewcommand\thetable{S.\arabic{table}} 


%


\section{I. System setup}
\label{sec:supp system setup}

\begin{figure}
\centering
\includegraphics{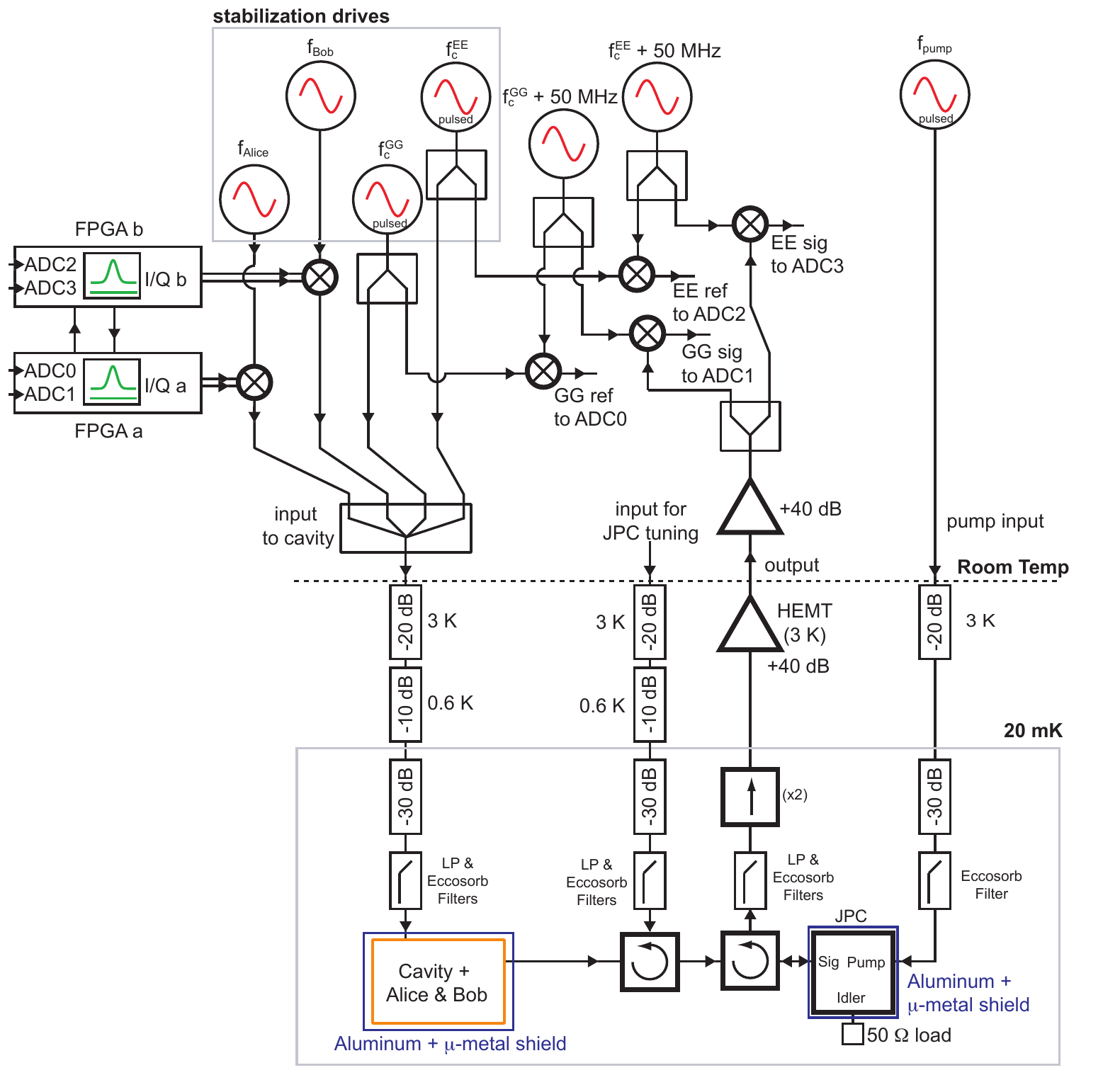}
\caption{\label{fig:exp_setup}}
\end{figure}

The cavity-qubit system was housed in an Oxford Triton 200 dilution refrigerator, at a base temperature below 20 mK. The setup of the experiment is described in detail in Fig.~\ref{fig:exp_setup}. The key feature is that both data acquisition and arbitrary waveform/digital marker generation were accomplished by two FPGA boards programmed with a customized logic.

The cavity frequency is $f_{gg} = 7.5$~GHz when both qubits are in the ground state and its linewidth $\kappa/2\pi$ is $2$~MHz. The qubit frequencies with no cavity photons are $\omega_{\text{Alice}}^0/2\pi = 4.87$~GHz and $\omega_{\text{Bob}}^0/2\pi = 6.18$~GHz, for Alice and Bob respectively. The two dispersive shifts are $\chi_{\text{Alice}}/2\pi = 5$~MHz, $\chi_{\text{Bob}}/2\pi = 4.5$~MHz. The anharmonicity for the two qubits are $c_{\text{Alice}}/2\pi= 212$~MHz, $c_{\text{Bob}}/2\pi= 209$~MHz. Alice (Bob) has a $T_1$ of $60$~$\mu$s ($18$~$\mu$s), $T_{2,\text{Ramsey}}$ of $9$~$\mu$s ($10$~$\mu$s) and excited state population in $\ket{eg}$ and $\ket{ge}$ of 5\% each.

As mentioned in the main text, the fast dispersive single-shot readout used in the experiment is made possible by a nearly quantum-limited phase-preserving measurement chain based on the Josephson Parametric Converter(JPC). The JPC was operated with a gain of $20$~dB and bandwidth of 6.2 MHz, with the frequency for maximum gain centered between the two readout frequencies, $f_{gg}$ and $f_{ee}$. The gain and noise visibility ratio at both $f_{gg}$ and $f_{ee}$ were about 17 dB and 6 dB, respectively. 

\section{II. Measurement strength and duration calibration for MB}
\refstepcounter{section}

\begin{figure}
\centering
\includegraphics{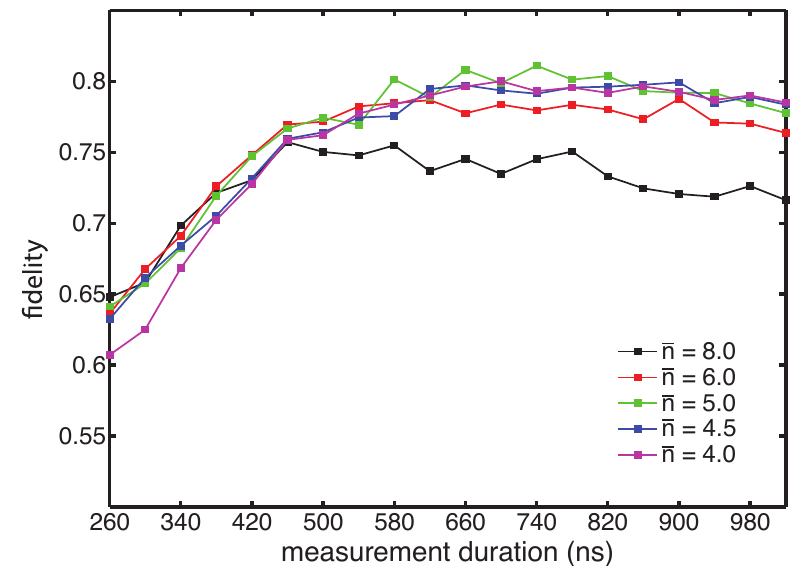}
\caption{\label{fig:measurement_calibration} Experimental calibration of measurement strength and duration for MB. Fidelity to the state $\ket{{\phi}_{+}}=\frac{1}{\sqrt{2}}(\ket{ge}+\ket{eg})$ after preparing the qubits in a maximally superposed state followed by a quasi-parity measurement with post-selection. The fidelity is plotted as a function of measurement duration and shown for a set of measurement strengths. The optimal values chosen for the experiment are 660 ns and $\bar{n}=4.5$.}
\end{figure}

\label{sec:msmt time}
The quasi-parity measurement strength and duration were optimized in order to maximize the fidelity of MB. This optimization was done by maximizing the fidelity of the Bell state created in a calibration experiment, similar to Ref.~\onlinecite{Riste2013Nature}. The qubits are prepared in ground states (by post-selection) and then two $\pi/2$ pulses, are applied to Alice and Bob, producing the state  $\ket{{\psi}}=\frac{1}{2}(\ket{gg}+\ket{ee}+\ket{ge}+\ket{eg})$. The quasi-parity measurement, consisting of the two cavity drives on $f_{ee}$ and $f_{gg}$ respectively, projects the qubits into one of the two even states or entangles the qubits into a Bell state with odd parity. We varied the duration of this parity measurement and its strength in terms of photon number (set to be identical) for each readout frequency to find the parameters that maximize the fidelity of the entangled state to the closest Bell state (Fig.~\ref{fig:measurement_calibration}). The Bell state fidelity would ideally increase and asymptotically approach one with increasing measurement time as the parity measurement better distinguishes the odd Bell state from the even states. On the other hand, at long measurement duration, the coherence of the entangled state decreases due to both natural and measurement-induced dephasing, the latter of which is caused by the $\chi$ mismatch between the qubits and is proportional to the average number of photons used for the measurement \cite{Tornberg2010}. Therefore, there is an optimal measurement strength and duration. For our experiment, $\bar{n}=4.5$ for each readout frequency and a measurement duration of $660$~ns are found to be close to the optimal values and are chosen to attain a Bell state fidelity of $80$\%. 

The value of $80$\% sets the upper bound  on the fidelity that we should expect for heralding MB. In the actual MB experiment, an extra $310$~ns delay was introduced after the quasi-parity measurement in a correction step, which does not occur in the sequence described in this section for optimizing the parity measurement parameters. This extra delay was required to accommodate the feedback latency in MB. The conditioned fidelity we obtained for heralded MB is about 6\% lower.

\section{III. Measurement outcomes distribution for DD and MB}
\refstepcounter{section}

\begin{figure}
\centering
\includegraphics{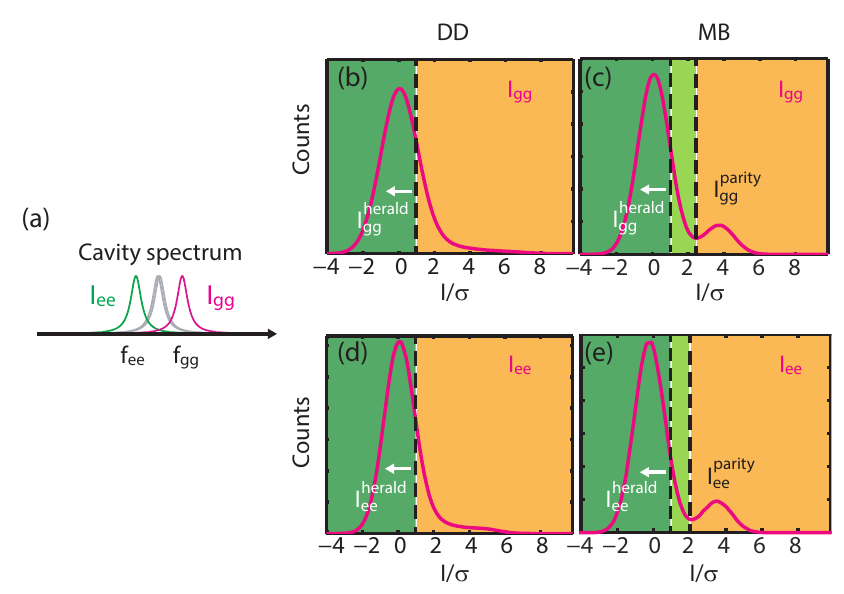}
\caption{\label{fig:thresh_hist}(a) The cavity outputs at $f_{gg}$ and $f_{ee}$ are integrated to obtain measurement outcomes $I_{gg}$ and $I_{ee}$, respectively. (b),(c), (d), (e) Histograms of the measurement outcomes $I_{gg}$ and $I_{ee}$ for DD and MB, respectively. The leftmost dashed line (labeled $I_{gg}^{herald}$  and $I_{ee}^{herald}$) indicates a particular choice of threshold for heralding the measurement outcomes. Counts to the left of both thresholds are declared success. The right dashed line ($I_{gg}^{parity}$ and $I_{ee}^{parity}$) indicates the threshold for quasi-parity measurement in MB.}
\end{figure}

\label{sec:thresholds}
The cavity outputs at $f_{gg}$ and $f_{ee}$ for both DD and MB can be used to monitor the state of the qubits during stabilization (Fig.~\ref{fig:thresh_hist}a). Histograms of the measurement outcomes $I_{gg}$ and $I_{ee}$, recorded by integrating the cavity output at $f_{gg}$ and $f_{ee}$, respectively, are shown in Fig.~\ref{fig:thresh_hist}b,c,d,e for DD and MB respectively. In DD, the cavity output signals are captured while all the CW drives are still on, i.e., the qubits are being driven while their states are being monitored, whereas in MB the outputs are a result of the quasi-parity measurements which occur after the qubit pulses and thus when the qubits are not driven. We observe that the measurement outcome distribution of DD lacks the separation seen in MB which has a clear parity separatrix $\{I_{gg}^{parity},I_{ee}^{parity}\}$. This feature also appears in numerical simulations of DD by the stochastic master equation \cite{Matti2015_}. The state estimation used in both DD and MB uses the ``box car'' filtering \cite{Gambetta2007} which simply sums up the recorded cavity output signals over time to obtain the measurement outcomes. This method, while appropriate for MB, is not suited for DD since in the latter, the qubits are undergoing actively-driven dynamics when the measurement is taking place. A more advanced filter, such as a non-linear quantum filter can be designed from either first-principles or machine learning \cite{Magesan2015} in the future to improve the state discrimination accuracy in DD.

The measurement outcomes to the left of both the $I_{gg}^{herald}$ (shown in figure) and $I_{ee}^{herald}$  thresholds are much less likely to come from even states than those to the right. Therefore the experiment runs with these outcomes are selected for state tomography, giving the results plotted as a color map in Fig.~3 (main text). Moving the threshold further to the left increases the stringency of the threshold as fewer measurement outcomes are included. The success probability for each threshold choice (plotted as contours in Fig.~3) is calculated by the ratio of included outcomes to the total number of experiment runs.

\section{IV. Steady state model for DD and MB}
\refstepcounter{section}

\label{sec:supp model}

The steady state behavior of both DD and MB is simulated by a Lindblad master equation, given by
\begin{equation}
\frac{d\rho(t)}{dt}=-\frac{i}{\hbar}[H(t),\rho(t)]+\kappa D[a]\rho(t)+\sum\limits_{j=A,B}\left(\frac{1}{T_{\downarrow}^j}D[\sigma_-^j]\rho(t)+\frac{1}{T_{\uparrow}^j}D[\sigma_+^j]\rho(t)+\frac{1}{2T_{\phi}^j}D[\sigma_z^j]\rho(t)\right)
 \end{equation}
$D$ is the Lindblad super-operator, defined for an operator $O$ as $D[O]\rho=O\rho O^{\dagger}-(1/2)O^{\dagger}O\rho-(1/2)\rho O^{\dagger}O$. The pure dephasing rate for Alice and Bob, respectively, is given by $1/T_{\phi}^{A,B}=1/T_2^{A,B}-1/2T_1^{A,B}$, where $1/T_{1}^{A,B}=1/T_{\downarrow}^{A,B}+1/T_{\uparrow}^{A,B}$.

The Hamiltonian $H(t)$ is treated differently in DD and MB. For DD, the Hamiltonian is described in detail in the theory proposal \cite{Leghtas2013} and parameters in the Hamiltonian, such as the cavity and qubit drive amplitudes, are swept in simulation to find the optimal values. The optimal value for the cavity drive amplitude is found to be $\kappa\sqrt{\bar{n}}/2$ with $\bar{n}=4.0$, and $\kappa/2$ for the qubit drive amplitudes at both zero-photon and n-photon qubit frequencies. The DD simulation predicts a characteristic time constant of 1 $\mu$s and a steady state fidelity of 76\% (accounting for the delay between stabilization and state tomography to allow remaining cavity photons to decay).


For MB, a correction step is broken into four segments for effectively piecewise master equation simulation. The first part contains the conditional Rabi pulses which are simulated as perfect instantaneous unitary operations on the qubits. The second part is the decay during the pulses (154 ns total). The Hamiltonian during this part is just the the dispersive interaction between the qubits and the cavity, $H(t) = H_{disp}=\left(\chi_A\sigma_z^A /2 + \chi_B \sigma_z^B /2 \right)a^{\dagger}a$, in the rotating frame of the two qubits ($\omega_A^0$, $\omega_B^0$) and the cavity mode ($(\omega_c^{gg}+\omega_c^{ee})/2$). The third part is the quasi-parity measurement during which the cavity drives at the $f_{gg}$ and $f_{ee}$ resonances are on and the Hamiltonian is given by $H(t)= H_{disp}+2\epsilon_c \text{cos}\left((\chi_A+\chi_B)t/2\right)(a+a^{\dagger})$, where $\epsilon_c$ is the amplitude of the cavity drive (660 ns total). The last part is the remainder of the correction step, incurred by the latency of the feedback during which all drives are off and the qubit-cavity system is in free-decay. The dynamics in this part is again simulated by the dispersive interaction, $H(t) = H_{disp}$ (686 ns total). For the piecewise master equation simulation of a complete correction step as four segments, the density matrix at the end of a segment is used as the initial density matrix for the next segment. 

\begin{figure}
\centering
\includegraphics{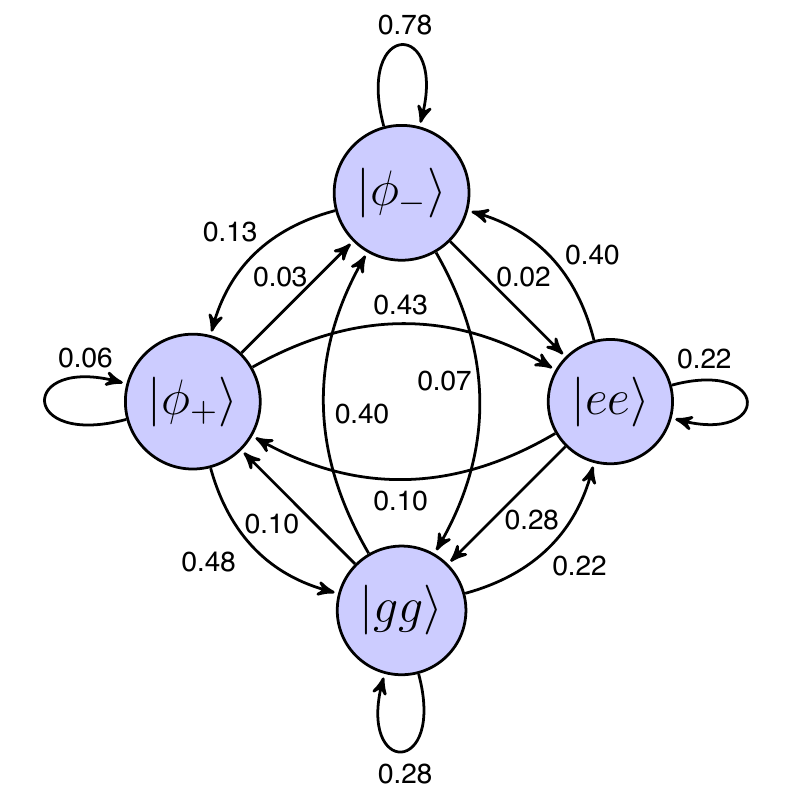}
\caption{\label{fig:mb_markov} Markov model of a correction step in MB. Transition between any two nodes is possible and is represented as a directional edge. The 16 possible transitions make up the 16 matrix elements in the stochastic transition matrix, $\mathcal{T}$, corresponding to a correction step in MB. Numbers next to the edges represent the elements of $\mathcal{T}$ calculated for the current experiment parameters by a master equation simulation.}
\end{figure}

Since in MB, the state at the end of a correction step depends only on the initial state at the beginning of the step, we can model the MB scheme as a Markov chain (Fig.~\ref{fig:mb_markov}). In the rest of this section, we show how we derive the transition matrix that describes this Markov chain. 

As discussed in the main text, we can describe the qubits by the density matrix $\rho= \pi_{-}\ket{{\phi}_{-}}\bra{{\phi}_{-}} + \pi_{+}\ket{{\phi}_{+}}\bra{{\phi}_{+}} + \pi_{gg}\ket{gg}\bra{gg} + \pi_{ee}\ket{ee}\bra{ee}$. Therefore, in terms of probability distributions in the four basis states, the qubits's state, $\tilde{S}$, can be represented by a vector, 
\begin{equation}
\tilde{S} = \begin{pmatrix}\pi_{-}\\ \pi_{+}\\ \pi_{gg}\\ \pi_{ee} \end{pmatrix}.
\end{equation}
If the qubits are prepared in $\ket{{\phi}_{-}}$, that is $\tilde{S}_{\phi_{-}}^{(i)}=(1,0,0,0)^{\text{T}}$, we can calculate $\tilde{S}_{\phi_{-}}^{(f)}$ after a correction step by applying the master equation simulation method described above. We need to consider the two possible cases where the conditional unitary applied is $U_O$ or $U_E$, respectively. The $\tilde{S}_{\phi_{-}}^{(f)}$ is a weighted average of the two cases.
\begin{equation}
\tilde{S}_{\phi_{-}}^{(f)}=(1-\epsilon_{E|O})*\tilde{S}_{{\phi_{-}}|U=U_O}^{(f)}+\epsilon_{E|O}*\tilde{S}_{{\phi_{-}}|U=U_E}^{(f)}
\end{equation}
$\epsilon_{E|O}$ and $\epsilon_{O|E}$ are the quasi-parity measurement infidelities due to limited measurement efficiency, introduced in the main text. In a similar manner, we can obtain $\tilde{S}_{\phi_{+}}^{(f)}$. In the case of an even initial state, for example, $\tilde{S}_{gg}^{(i)}=(0,0,1,0)^{\text{T}}$, we have
\begin{equation}
\tilde{S}_{gg}^{(f)}=(1-\epsilon_{O|E})*\tilde{S}_{gg|U=U_E}^{(f)}+\epsilon_{O|E}*\tilde{S}_{gg|U=U_O}^{(f)}
\end{equation}
And similarly for $\tilde{S}_{ee}^{(f)}$. 

Given $\tilde{S}_{\phi_{-}}^{(f)}$, $\tilde{S}_{\phi_{+}}^{(f)}$, $\tilde{S}_{gg}^{(f)}$ and $\tilde{S}_{ee}^{(f)}$, we can construct the transition matrix $\mathcal{T}$ of a correction step, 
\begin{equation}
\mathcal{T}=(\tilde{S}_{\phi_{-}}^{(f)}, \tilde{S}_{\phi_{+}}^{(f)}, \tilde{S}_{gg}^{(f)}, \tilde{S}_{ee}^{(f)})
\end{equation}
where the $\tilde{S}^{(f)}$'s are the columns of the 4 by 4 matrix. Now applying this transition matrix on any arbitrary initial state gives the final state after a correction step, 
\begin{equation}
\tilde{S}^{(f)}=\mathcal{T}\tilde{S}^{(i)}
\end{equation}

The transition matrix $\mathcal{T}$  is also called the stochastic matrix, with the property that each column sums to 1. One of $\mathcal{T}$'s eigenvalues is guaranteed to be 1 and the corresponding eigenvector, $\tilde{S}_{\infty}$, is the steady state of the Markov chain.  It can easily be shown that for any arbitrary initial state, $\tilde{S}^{(i)}$
\begin{equation}
\lim_{k\to\infty} \mathcal{T}^{k}\tilde{S}^{(i)} = \tilde{S}_{\infty}
\end{equation}
For the given experimental parameters in MB, the $\mathcal{T}$ matrix is displayed in Fig.~\ref{fig:mb_markov} and we find the steady state eigenvector to be
\begin{equation}
\tilde{S}_{\infty}=\begin{pmatrix}0.58\\ 0.11\\ 0.18\\ 0.13 \end{pmatrix}
\end{equation}
Thus, the Markov model predicts a steady state fidelity of 58\% to $\ket{{\phi}_{-}}$. Taking into account of the duration of a correction step (1.5 $\mu$s), we can also calculate the characteristic time constant of the MB scheme from the model, which gives 1.4 $\mu$s. Both values agree very well with experimental results. 

We can calculate the expected fidelity when some of the experimental parameters are improved in the near future. If the measurement efficiency is improved from 30\% to the current state-of-the-art value of 60\%, the measurement duration can be reduced by half while maintaining the quasi-parity measurement infidelities\cite{Hatridge2013}. The instrument and FPGA latencies incurred in the experiment can also be reduced by 100 ns, in the latest hardware setup and FPGA logic design in operation while this article was being prepared.  The measurement duration and control latency reduction can shorten the correction step length to 1 $\mu$s, which can improve the steady state fidelity to 66\%. Furthermore, if the coherence times are also improved to the state-of-the-art values in the hundred of microseconds for superconducting qubits, both DD and MB fidelities in the ``fixed time'' protocol can be above 85\%, limited by the $\chi$ mismatch (assumed to be 10\%, as in the current experiment). The prospects of both DD and MB schemes are summarized in Tab.~\ref{tab:fidelity_prospects}

\begin{table}
\begin{tabular}{l | l  P{2cm}  P{2cm}}
	\hline
	\hline
	\multicolumn{2}{c}{} & \multicolumn{2}{c}{Steady state fidelity} \\
	\multicolumn{2}{c}{} & DD & MB \\ \hline 
	Experiment & current parameters & 76\% & 57\% \\ \hline
	\multirow{2}{*}{Simulation} & $\eta=0.6$, latency\,$=586$~ns & 76\% & 66\%  \\ 
	& $T_1$,$T_2 = 100$~$\mu$s & 86\% & 86\% \\ \hline
	\hline
\end{tabular}
\caption{\label{tab:fidelity_prospects} Listing of the steady state fidelities of the ``fixed time'' protocol for DD and MB schemes for the current experiment (see detailed parameters in Supp.~Sec. ~\ref{sec:supp system setup}) and simulation with improved parameters, assuming $\chi$ values as in the current experiment. The prospective values in the last row also take into account the parameter changes in the second row. Note that measurement efficiency and control latency change do not affect DD. }
\end{table}

\section{V. Simulation of real-time heralding by nested feedback protocol for DD and MB}
\refstepcounter{section}

\begin{figure}
\centering
\includegraphics{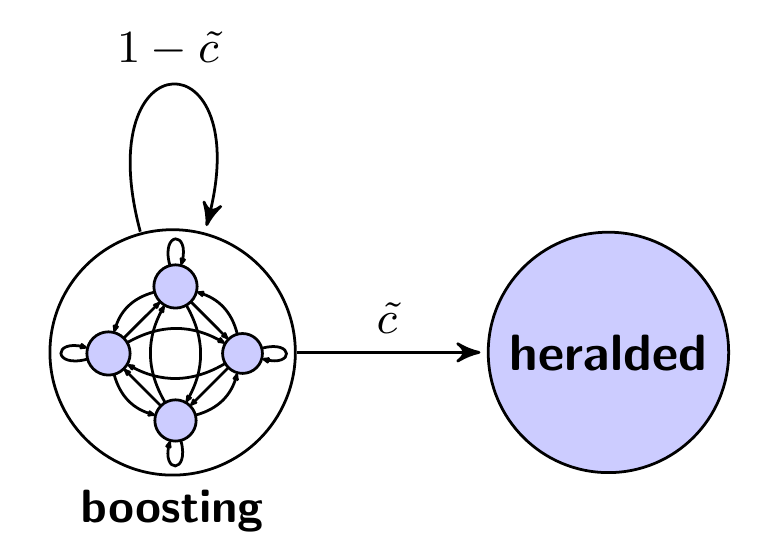}
\caption{\label{fig:nfp_markov} Markov model of the NFP (nested feedback protocol) for real-time heralding. This model is an extension of the model introduced for MB. During a boost attempt, transitions occur inside the ``boosting'' node for further stabilization. The edge leading from the ``boosting'' to the ``heralded'' node represents the real-time heralding that selects some fraction of trajectories by the threshold-dependent matrix $\tilde{c}$ (see text) at the end of a boost attempt. Those that are not selected ($\bm{1}-\tilde{c}$) enter into another boost attempt. }
\end{figure}

\label{sec:nfp_simulation}
The Markov chain model introduced in Sec. ~\ref{sec:supp model} can be extended to simulate NFP (nested feedback protocol). We can construct a ``nested'' Markov chain (Fig.~\ref{fig:nfp_markov}) for each boost attempt of NFP. At the outer-most level, there are two nodes. One node denotes the trajectories that have just been heralded as successful; the other node denotes the trajectories that require at least another boost attempt.  Building on the vector description established in the previous section, we represent the heralding of trajectories before a boost attempt by a diagonal matrix $\tilde{c}$, 

\begin{equation}
\tilde{c} = 
\begin{pmatrix}
c_{\phi_-}\\
&c_{\phi_+}\\
&&c_{gg}\\
&&&c_{ee}
\end{pmatrix}
\end{equation}

After stabilization of some pre-determined duration, if the qubits are (on average) in state $\tilde{S}^{(0)}$, then the average state of the qubits that are heralded is then given by,

\begin{equation}
\tilde{S}_{herald}^{(0)}=\frac{\tilde{c}\tilde{S}^{(0)}}{||\tilde{c}\tilde{S}^{(0)}||_1}\,,
\end{equation}
where $||\cdot||_1$ is the $L_1$ norm of the vector (the sum of the entries in the vector). The normalization is required since heralding selects only a subset of the trajectories, i.e., the matrix, $\tilde{c}$, does not preserve the norm of $\tilde{S}$.

Each entry on the diagonal of $\tilde{c}$ gives the fraction of trajectories that get heralded (selected) from a particular state. Their values depend on the particular heralding thresholds used, $I_{gg}^{herald}$ and $I_{ee}^{herald}$ (Supp.~Sec. ~\ref{sec:thresholds}). This matrix can be determined phenomenologically by the post-selection experiment shown in Fig.~3 (main text). From the experiment, we can find the average state without any conditioning (no heralding), $\tilde{S}^{(0)} = (\pi_{-}^{(0)},\pi_{+}^{(0)}, \pi_{gg}^{(0)},\pi_{ee}^{(0)})^{\text{T}}$ and the average state of the heralded trajectories, $\tilde{S}_{herald}^{(0)} = (\pi_{-}',\pi_{+}', \pi_{gg}',\pi_{ee}')^{\text{T}}$. Given the success probability $P_s$ of using the thresholds (the white dashed contour line of Fig.~3b,d in the main text), the diagonal elements can be calculated as,

\begin{equation}
c_{\phi_-}=\frac{\pi_{-}'P_s}{\pi_{-}^{(0)}}\,, \quad
c_{\phi_+}=\frac{\pi_{+}'P_s}{\pi_{+}^{(0)}}\,, \quad
c_{gg}=\frac{\pi_{gg}'P_s}{\pi_{gg}^{(0)}}\,, \quad
c_{ee}=\frac{\pi_{ee}'P_s}{\pi_{ee}^{(0)}} 
\end{equation}

For the specific heralding thresholds used in the experiment (represented by the black squares in Fig.~3b,d in the main text), $\tilde{c}_{\text{DD}}$ and $\tilde{c}_{\text{MB}}$ are explicitly given by,

\begin{equation}
\tilde{c}_{\text{DD}} = 
\begin{pmatrix}
0.26\\
&0.20\\
&&0.19\\
&&&0.18
\end{pmatrix}, \quad
\tilde{c}_{\text{MB}} = 
\begin{pmatrix}
0.68\\
&0.69\\
&&0.19\\
&&&0.10
\end{pmatrix}
\end{equation}

Ideally for $\tilde{c}$, only $c_{\phi_-}$ should be non-zero. But in practice, since we cannot distinguish $\ket{\phi_-}$ and  $\ket{\phi_+}$,  $c_{\phi_+}$ is comparable to $c_{\phi_-}$. Furthermore, for both DD and MB, $c_{gg}$ and $c_{ee}$ are also non-neglibile. In DD, this is predominantly due to the lack of separation between the even and odd measurement outcomes as discussed in Supp.~Sec.~\ref{sec:thresholds}. In MB, the qubits can jump during the delay between the completion of the quasi-parity measurement and the end of a correction step due to $T_1$ events. Thus for MB, trajectories that are heralded by very stringent thresholds still have a non-zero probability of being in the even parity states.

Given the heralding matrices $\tilde{c}_{\text{DD}}$ and $\tilde{c}_{\text{MB}}$, we can now calculate the average state of the trajectories that are not heralded and thus require a boost attempt as 

\begin{equation}
\tilde{s}_{boost}^{(0)} = (\bm{I}-\tilde{c})\tilde{S}^{(0)} \,,\quad
\tilde{S}_{boost}^{(0)}= \frac{\tilde{s}_{boost}^{(0)}}{||\cdot||_1}
\end{equation}
where we introduce the lowercase $\tilde{s}_{boost}^{(0)}$ as the unnormalized population distribution vector. $||\cdot||_1$ denotes the $L_1$ norm of the numerator.

After this boost attempt, the qubits are in state $\tilde{S}^{(1)}$, 

\begin{equation}
\tilde{S}^{(1)}= \frac{\mathcal{T}\tilde{s}_{boost}^{(0)}}{||\cdot||_1} = \frac{\mathcal{T}(\bm{I}-\tilde{c})\tilde{S}^{(0)}}{||\cdot||_1}
\end{equation}
where $\mathcal{T}$ is the stochastic transition matrix that models the stabilization during a boosting attempt. In Supp.~Sec. ~\ref{sec:supp model}, we have already found $\mathcal{T}$ for MB. By the same method, we can also derive the effective transition matrix of a boost attempt for DD. The calculation of $\mathcal{T}$ for DD is an approximation:  due to the continuous cavity drives, the state of the qubits at the beginning of a boost attempt is entangled with a qubit-state dependent cavity state, which we approximate unconditionally by the average steady-state cavity state in DD. Nonetheless, as we shall show, the model still produces a quantitative behavior that agrees very well with the experimental results. From the above equations, it is easy to show that the average qubits state of the heralded trajectories after $k$ boost attempts is given by

\begin{equation}
\tilde{S}_{herald}^{(k)}= \frac{\tilde{c} \left(\mathcal{T}(\bm{I}-\tilde{c})\right)^{k} \tilde{S}^{(0)}}{||\cdot||_1}
\end{equation}

In the case of a pre-stabilization of sufficient number of correction steps (or duration), $\tilde{S}^{(0)}$ is given by the steady-state, $\tilde{S}_{\infty}$, introduced in Supp.~Sec. ~\ref{sec:supp model}.

\begin{figure}
\centering
\includegraphics{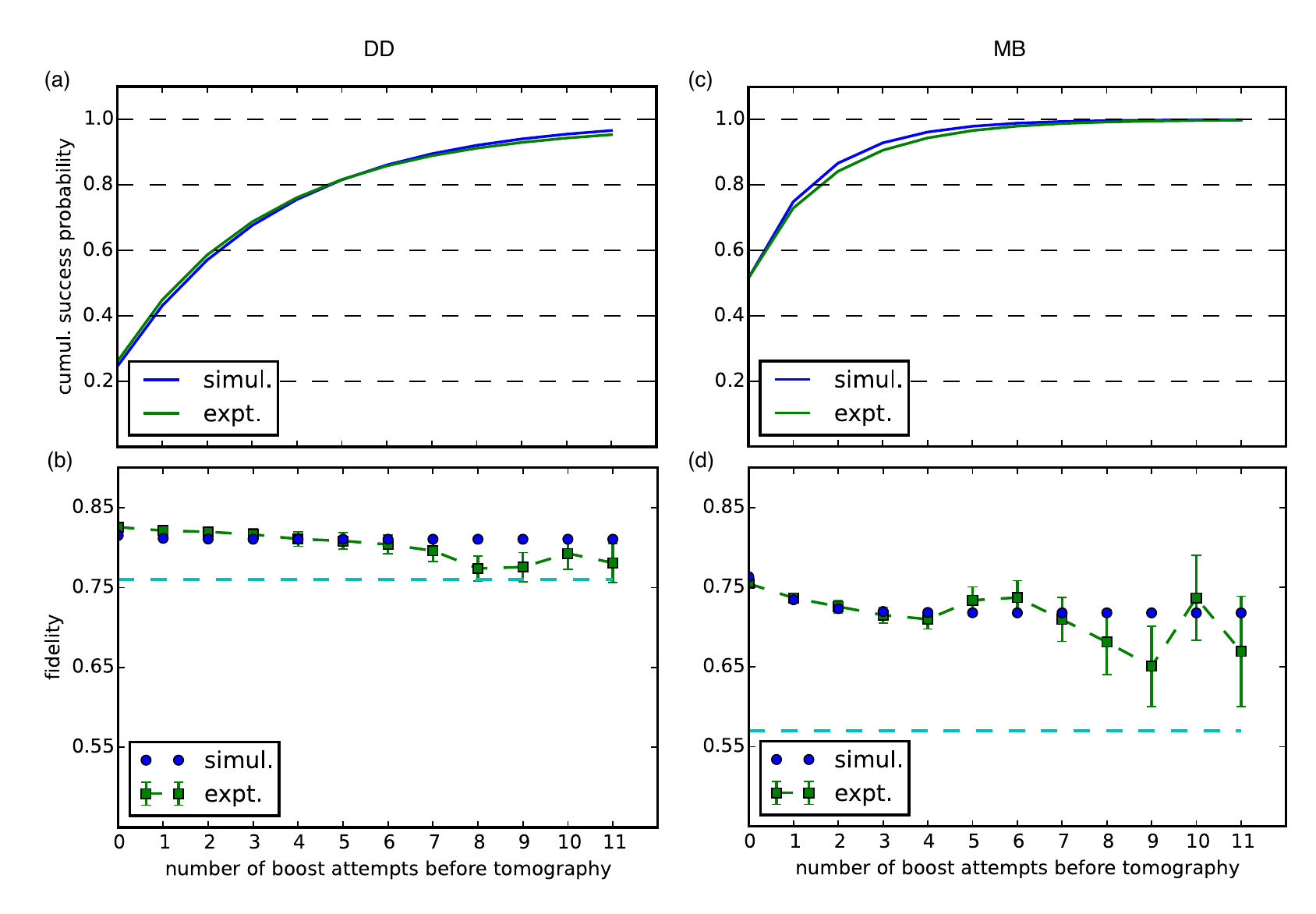}
\caption{\label{fig:nfp_sim_results} Simulation of real-time heralding by NFP. The left (right) panel presents the results for DD (MB). (a), (c) Cumulative success probability of having completed at most a given number of boost attempts before tomography for DD and MB, respectively. Green curve shows the experimental result as presented in the main text. Blue curve is the simulation result using the same experimental parameters. (b), (d) Fidelity to $\ket{\phi_-}$ for DD and MB, respectively. Cyan dashed line denotes the unconditioned steady state fidelity obtained in the experiment. Green squares (blue circles) show the corresponding fidelity as a function of the number of boost attempts during NFP obtained in the experiment (simulation).}
\end{figure}

For each of the $k$ boost attempts, the first entry in the vector $\tilde{S}_{herald}^{(k)}$, i.e., $\pi_{-}'$, gives the fidelity to $\ket{\phi_{-}}$. The  $L_1$ norm of the numerator in the expression for $\tilde{S}_{herald}^{(k)}$ gives the percentage of trajectories that have completed $k$ boost attempts. Summing the percentages over all values of $k$ from 0 to the maximum limit gives the overall success probability. In Fig.~\ref{fig:nfp_sim_results}, we show the results of the simulation using the model described here and compare them to the experimental results presented in Fig.~4 of the main text.  Furthermore, from the simulation, we find that with the realistic system parameter improvement as specified in Supp.~Sec. ~\ref{sec:supp model}, the fidelity of heralded trajectories can be 90\% and 95\% for DD and MB, respectively, with order unity success probability.

\section{VI. Measurement-induced AC stark shift and correction for MB}
\refstepcounter{section}

\begin{figure}
\centering
\includegraphics{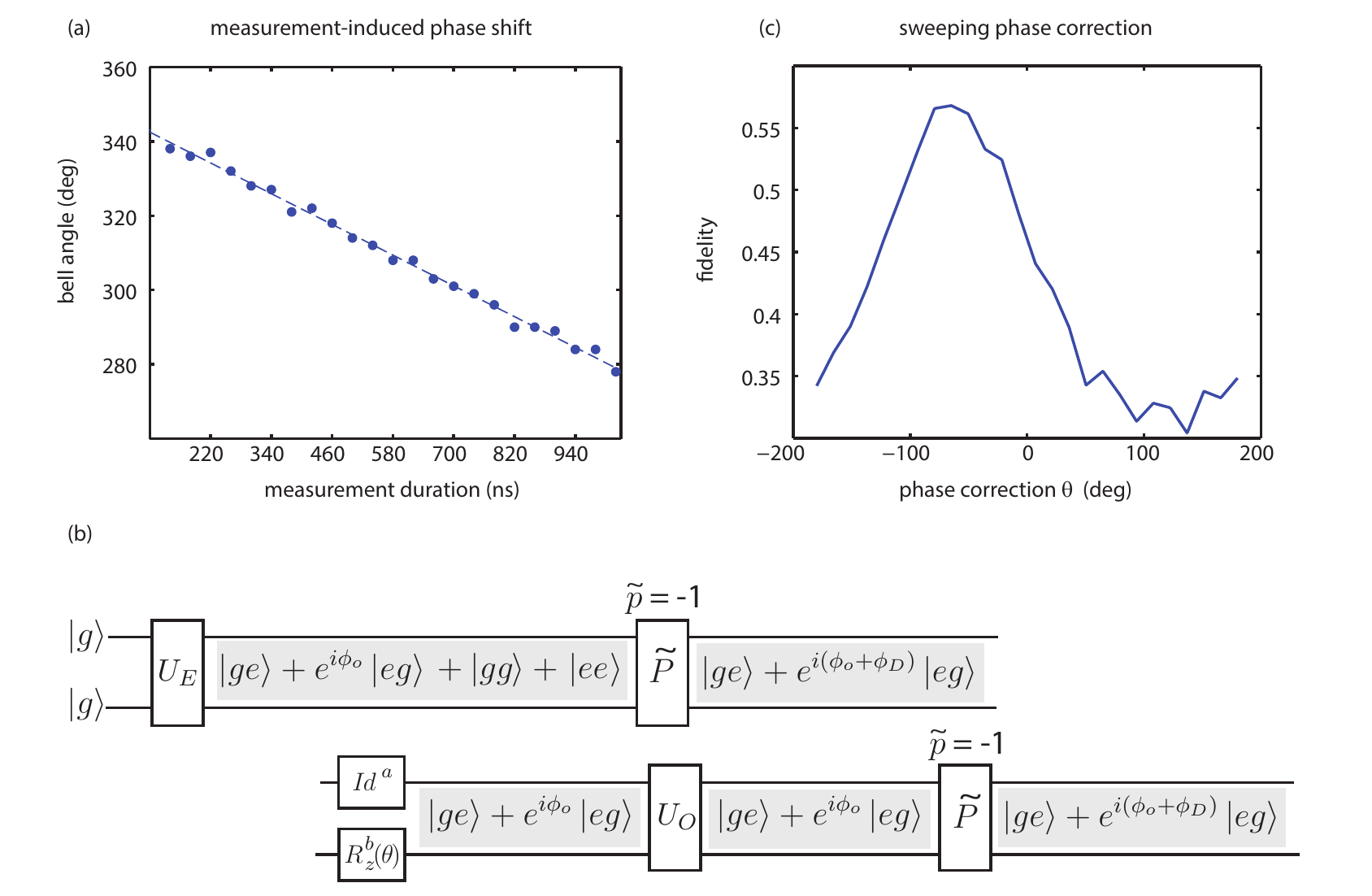}
\caption{\label{fig:measurement_dephasing} Correcting for deterministic measurement-induced AC Stark shift (a) Deterministic phase shift induced by measurement as a function of measurement duration. (b) One example of a sequence trajectory illustrating the phase correction $R_z^b(\theta)$ at work. See detailed explanation in accompanying text. (c) Fidelity of the steady state to the target Bell state as a function of the correction angle of the effective \textit{Z} gate on Bob. }
\end{figure}

\label{sec:AC stark shift}
The quasi-parity measurement induces a deterministic phase shift between the two qubits due to measurement-induced AC Stark shift \cite{Tornberg2010}. This is evidenced by examining the phase of the Bell state created in the measurement optimization experiment described in Supp.~Sec.~\ref{sec:msmt time}, in which we varied the measurement duration. As the measurement duration is increased, the Bell angle of the final Bell state changes linearly (Fig.~\ref{fig:measurement_dephasing}a). In order for MB to work, we need to account for the deterministic phase shift induced by the measurement. This correction is accomplished by a ``\textit{Z}'' rotation on Bob before the unitary $U_O$. Fig.~\ref{fig:measurement_dephasing}b gives one example of a sequence trajectory to illustrate how the correction works. With no loss of generality (and the reason will become clear soon in the discussion that follows), we construct $U_E=R_x^a(\pi/2)\otimes R_{-\phi_o}^b(\pi/2)$ and $U_O=R_x^a(\pi/2)\otimes R_{\phi_o}^b(\pi/2)$ (where $\phi_o$ = 0 corresponds to the X axis) such that  $\ket{\phi_{o}}=\ket{ge}+e^{i\phi_{o}}\ket{eg}$ is the eigenstate of $U_O$  and applying $U_E$ on the even states results in $\ket{\phi_{o}}$ with 50\% probability after the quasi-parity measurement. Suppose that the qubits are in the ground states, $U_E$ is applied and the subsequent quasi-parity measurement gives $\tilde{p} = -1$. During the quasi-parity measurement, a deterministic phase shift of $\phi_{D}$ is added. Since the measurement reports odd, the next conditional unitary is $U_O$. The \textit{Z} gate before $U_O$ undoes the phase shift and recovers the eigenstate which $U_O$ leaves unchanged. After another parity measurement, the qubits are in the state $\ket{ge}+e^{i(\phi_{o}+\phi_{D})}\ket{eg}$, with the phase shift added again. Consequently, we can see that the MB sequence actually stabilizes $\ket{ge}+e^{i(\phi_{o}+\phi_{D})}\ket{eg}$ state. In practice, for our experiment, the \textit{Z} rotation for Bob is constructed from a composite of \textit{X} and \textit{Y} rotations, such that $R_{z}^{b}(\theta)=R_{x}(\frac{\pi}{2})R_{y}(\theta)R_{x}(-\frac{\pi}{2})$, and the effective correction angle $\theta$ is swept (Fig.~\ref{fig:measurement_dephasing}c) to find the optimal value that cancels the deterministic phase shift and thus maximizes the fidelity. Furthermore, to make the target state of the stabilization  $\ket{\phi_{-}}$, the rotation axis of the pulses on Bob in $U_O$ and $U_E$ is chosen such that $\phi_{o}+\phi_{D}=\pi$. This correction for measurement-induced AC Stark shift is also done in the simulation for MB.

\section{VII. \textit{f} state measurement during NFP}
\refstepcounter{section}

\begin{figure}
\centering
\includegraphics{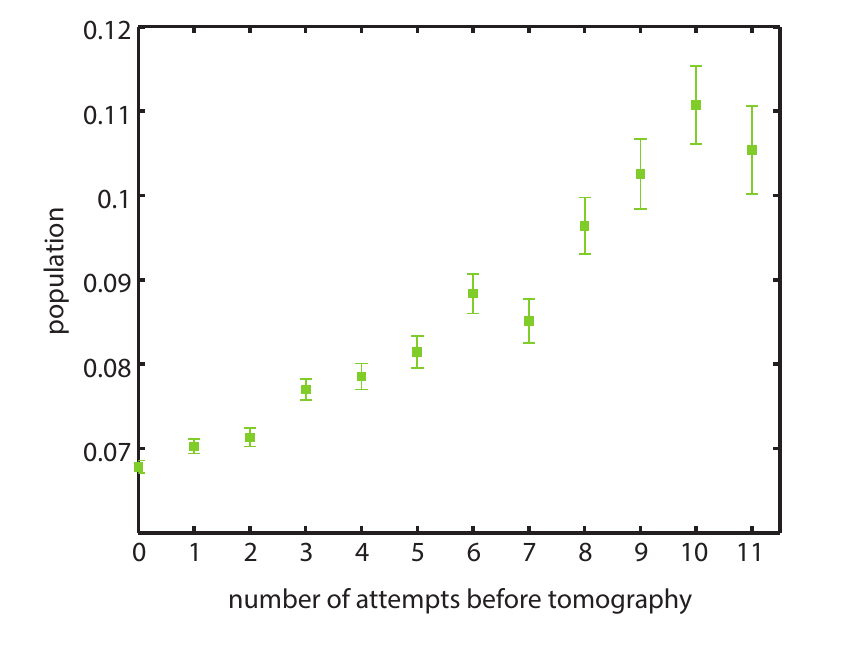}
\caption{\label{fig:f_pop} Experimental measurement of \textit{f} state population plotted as a function of the boost attempts number in NFP for DD.}
\end{figure}

\label{sec:f state}
While we have been treating our two qubits as purely two-level systems, in reality there are higher energy levels, in particular the second excited level is expected to play a non-negligible role in the dynamics. We find that the equilibrium qubit population not in the $\ket{gg}$ state was about $15$\%, and the $f$ state was also populated. To investigate whether the decrease of the fidelity in NFP as a function of boost attempts number is due to the role played by the $f$-state population, we measured the populations in $\ket{fg}$, $\ket{fe}$, $\ket{gf}$, $\ket{ef}$, $\ket{ff}$ after a given number of boost attempts in DD. The population in $\ket{fg}$ was measured by applying a $\pi$ pulse on Alice's \textit{e-f} transition, then a $\pi$ pulse on its \textit{e-g} transition, followed by a measurement of the population in $\ket{gg}$. The other $f$-state populations are similarly obtained. The sum of these 5 populations gives the total $f$-state population plotted in Fig.~\ref{fig:f_pop}, which indeed increases as a function of boost attempt number. It is therefore plausible that the decrease in fidelity with respect to number of boost attempts is due to the system being trapped in other levels outside the correction space. This problem, which is common in similar atomic physics experiments, could be addressed by additional drives.


%

\clearpage

%


\begin{thebibliography}{62}%
\makeatletter
\providecommand \@ifxundefined [1]{%
 \@ifx{#1\undefined}
}%
\providecommand \@ifnum [1]{%
 \ifnum #1\expandafter \@firstoftwo
 \else \expandafter \@secondoftwo
 \fi
}%
\providecommand \@ifx [1]{%
 \ifx #1\expandafter \@firstoftwo
 \else \expandafter \@secondoftwo
 \fi
}%
\providecommand \natexlab [1]{#1}%
\providecommand \enquote  [1]{``#1''}%
\providecommand \bibnamefont  [1]{#1}%
\providecommand \bibfnamefont [1]{#1}%
\providecommand \citenamefont [1]{#1}%
\providecommand \href@noop [0]{\@secondoftwo}%
\providecommand \href [0]{\begingroup \@sanitize@url \@href}%
\providecommand \@href[1]{\@@startlink{#1}\@@href}%
\providecommand \@@href[1]{\endgroup#1\@@endlink}%
\providecommand \@sanitize@url [0]{\catcode `\\12\catcode `\$12\catcode
  `\&12\catcode `\#12\catcode `\^12\catcode `\_12\catcode `\%12\relax}%
\providecommand \@@startlink[1]{}%
\providecommand \@@endlink[0]{}%
\providecommand \url  [0]{\begingroup\@sanitize@url \@url }%
\providecommand \@url [1]{\endgroup\@href {#1}{\urlprefix }}%
\providecommand \urlprefix  [0]{URL }%
\providecommand \Eprint [0]{\href }%
\providecommand \doibase [0]{http://dx.doi.org/}%
\providecommand \selectlanguage [0]{\@gobble}%
\providecommand \bibinfo  [0]{\@secondoftwo}%
\providecommand \bibfield  [0]{\@secondoftwo}%
\providecommand \translation [1]{[#1]}%
\providecommand \BibitemOpen [0]{}%
\providecommand \bibitemStop [0]{}%
\providecommand \bibitemNoStop [0]{.\EOS\space}%
\providecommand \EOS [0]{\spacefactor3000\relax}%
\providecommand \BibitemShut  [1]{\csname bibitem#1\endcsname}%
\let\auto@bib@innerbib\@empty
\bibitem [{\citenamefont {Nielsen}\ and\ \citenamefont
  {Chuang}(2004)}]{Nielsen2004}%
  \BibitemOpen
  \bibfield  {author} {\bibinfo {author} {\bibfnamefont {M.~A.}\ \bibnamefont
  {Nielsen}}\ and\ \bibinfo {author} {\bibfnamefont {I.~L.}\ \bibnamefont
  {Chuang}},\ }\href@noop {} {\emph {\bibinfo {title} {Quantum Computation and
  Quantum Information}}}\ (\bibinfo  {publisher} {Cambridge University Press},\
  \bibinfo {year} {2004})\BibitemShut {NoStop}%
\bibitem [{\citenamefont {Terhal}(2015)}]{Terhal2015quantum}%
  \BibitemOpen
  \bibfield  {author} {\bibinfo {author} {\bibfnamefont {B.~M.}\ \bibnamefont
  {Terhal}},\ }\href@noop {} {\bibfield  {journal} {\bibinfo  {journal}
  {Reviews of Modern Physics}\ }\textbf {\bibinfo {volume} {87}},\ \bibinfo
  {pages} {307} (\bibinfo {year} {2015})}\BibitemShut {NoStop}%
\bibitem [{\citenamefont {Kerckhoff}\ \emph {et~al.}(2010)\citenamefont
  {Kerckhoff}, \citenamefont {Nurdin}, \citenamefont {Pavlichin},\ and\
  \citenamefont {Mabuchi}}]{Kerckhoff2010}%
  \BibitemOpen
  \bibfield  {author} {\bibinfo {author} {\bibfnamefont {J.}~\bibnamefont
  {Kerckhoff}}, \bibinfo {author} {\bibfnamefont {H.~I.}\ \bibnamefont
  {Nurdin}}, \bibinfo {author} {\bibfnamefont {D.~S.}\ \bibnamefont
  {Pavlichin}}, \ and\ \bibinfo {author} {\bibfnamefont {H.}~\bibnamefont
  {Mabuchi}},\ }\href {\doibase 10.1103/PhysRevLett.105.040502} {\bibfield
  {journal} {\bibinfo  {journal} {Phys. Rev. Lett.}\ }\textbf {\bibinfo
  {volume} {105}},\ \bibinfo {pages} {040502} (\bibinfo {year}
  {2010})}\BibitemShut {NoStop}%
\bibitem [{\citenamefont {Fowler}\ \emph {et~al.}(2012)\citenamefont {Fowler},
  \citenamefont {Mariantoni}, \citenamefont {Martinis},\ and\ \citenamefont
  {Cleland}}]{Fowler2012}%
  \BibitemOpen
  \bibfield  {author} {\bibinfo {author} {\bibfnamefont {A.~G.}\ \bibnamefont
  {Fowler}}, \bibinfo {author} {\bibfnamefont {M.}~\bibnamefont {Mariantoni}},
  \bibinfo {author} {\bibfnamefont {J.~M.}\ \bibnamefont {Martinis}}, \ and\
  \bibinfo {author} {\bibfnamefont {A.~N.}\ \bibnamefont {Cleland}},\ }\href
  {\doibase 10.1103/PhysRevA.86.032324} {\bibfield  {journal} {\bibinfo
  {journal} {Phys. Rev. A}\ }\textbf {\bibinfo {volume} {86}},\ \bibinfo
  {pages} {032324} (\bibinfo {year} {2012})}\BibitemShut {NoStop}%
\bibitem [{\citenamefont {Fujii}\ \emph {et~al.}(2014)\citenamefont {Fujii},
  \citenamefont {Negoro}, \citenamefont {Imoto},\ and\ \citenamefont
  {Kitagawa}}]{Fujii2014}%
  \BibitemOpen
  \bibfield  {author} {\bibinfo {author} {\bibfnamefont {K.}~\bibnamefont
  {Fujii}}, \bibinfo {author} {\bibfnamefont {M.}~\bibnamefont {Negoro}},
  \bibinfo {author} {\bibfnamefont {N.}~\bibnamefont {Imoto}}, \ and\ \bibinfo
  {author} {\bibfnamefont {M.}~\bibnamefont {Kitagawa}},\ }\href {\doibase
  10.1103/PhysRevX.4.041039} {\bibfield  {journal} {\bibinfo  {journal} {Phys.
  Rev. X}\ }\textbf {\bibinfo {volume} {4}},\ \bibinfo {pages} {041039}
  (\bibinfo {year} {2014})}\BibitemShut {NoStop}%
\bibitem [{\citenamefont {Wiseman}\ and\ \citenamefont
  {Milburn}(2009)}]{Wiseman2009}%
  \BibitemOpen
  \bibfield  {author} {\bibinfo {author} {\bibfnamefont {H.~M.}\ \bibnamefont
  {Wiseman}}\ and\ \bibinfo {author} {\bibfnamefont {G.~J.}\ \bibnamefont
  {Milburn}},\ }\href@noop {} {\emph {\bibinfo {title} {Quantum Measurement and
  Control}}}\ (\bibinfo  {publisher} {Cambridge University Press},\ \bibinfo
  {year} {2009})\BibitemShut {NoStop}%
\bibitem [{\citenamefont {Sayrin}\ \emph {et~al.}(2011)\citenamefont {Sayrin},
  \citenamefont {Dotsenko}, \citenamefont {Zhou}, \citenamefont {Peaudecerf},
  \citenamefont {Rybarczyk}, \citenamefont {Gleyzes}, \citenamefont {Rouchon},
  \citenamefont {Mirrahimi}, \citenamefont {Amini}, \citenamefont {Brune},
  \citenamefont {Raimond},\ and\ \citenamefont {Haroche}}]{Haroche2011Nature}%
  \BibitemOpen
  \bibfield  {author} {\bibinfo {author} {\bibfnamefont {C.}~\bibnamefont
  {Sayrin}}, \bibinfo {author} {\bibfnamefont {I.}~\bibnamefont {Dotsenko}},
  \bibinfo {author} {\bibfnamefont {X.}~\bibnamefont {Zhou}}, \bibinfo {author}
  {\bibfnamefont {B.}~\bibnamefont {Peaudecerf}}, \bibinfo {author}
  {\bibfnamefont {T.}~\bibnamefont {Rybarczyk}}, \bibinfo {author}
  {\bibfnamefont {S.}~\bibnamefont {Gleyzes}}, \bibinfo {author} {\bibfnamefont
  {P.}~\bibnamefont {Rouchon}}, \bibinfo {author} {\bibfnamefont
  {M.}~\bibnamefont {Mirrahimi}}, \bibinfo {author} {\bibfnamefont
  {H.}~\bibnamefont {Amini}}, \bibinfo {author} {\bibfnamefont
  {M.}~\bibnamefont {Brune}}, \bibinfo {author} {\bibfnamefont {J.-M.}\
  \bibnamefont {Raimond}}, \ and\ \bibinfo {author} {\bibfnamefont
  {S.}~\bibnamefont {Haroche}},\ }\href {http://dx.doi.org/10.1038/nature10376}
  {\bibfield  {journal} {\bibinfo  {journal} {Nature}\ }\textbf {\bibinfo
  {volume} {477}},\ \bibinfo {pages} {73} (\bibinfo {year} {2011})}\BibitemShut
  {NoStop}%
\bibitem [{\citenamefont {Chiaverini}\ \emph {et~al.}(2004)\citenamefont
  {Chiaverini}, \citenamefont {Leibfried}, \citenamefont {Schaetz},
  \citenamefont {Barrett}, \citenamefont {Blakestad}, \citenamefont {Britton},
  \citenamefont {Itano}, \citenamefont {Jost}, \citenamefont {Knill},
  \citenamefont {Langer} \emph {et~al.}}]{Chiaverini2004}%
  \BibitemOpen
  \bibfield  {author} {\bibinfo {author} {\bibfnamefont {J.}~\bibnamefont
  {Chiaverini}}, \bibinfo {author} {\bibfnamefont {D.}~\bibnamefont
  {Leibfried}}, \bibinfo {author} {\bibfnamefont {T.}~\bibnamefont {Schaetz}},
  \bibinfo {author} {\bibfnamefont {M.}~\bibnamefont {Barrett}}, \bibinfo
  {author} {\bibfnamefont {R.}~\bibnamefont {Blakestad}}, \bibinfo {author}
  {\bibfnamefont {J.}~\bibnamefont {Britton}}, \bibinfo {author} {\bibfnamefont
  {W.}~\bibnamefont {Itano}}, \bibinfo {author} {\bibfnamefont
  {J.}~\bibnamefont {Jost}}, \bibinfo {author} {\bibfnamefont {E.}~\bibnamefont
  {Knill}}, \bibinfo {author} {\bibfnamefont {C.}~\bibnamefont {Langer}},
  \emph {et~al.},\ }\href@noop {} {\bibfield  {journal} {\bibinfo  {journal}
  {Nature}\ }\textbf {\bibinfo {volume} {432}},\ \bibinfo {pages} {602}
  (\bibinfo {year} {2004})}\BibitemShut {NoStop}%
\bibitem [{\citenamefont {Nigg}\ \emph {et~al.}(2014)\citenamefont {Nigg},
  \citenamefont {M{\"u}ller}, \citenamefont {Martinez}, \citenamefont
  {Schindler}, \citenamefont {Hennrich}, \citenamefont {Monz}, \citenamefont
  {Martin-Delgado},\ and\ \citenamefont {Blatt}}]{Nigg2014quantum}%
  \BibitemOpen
  \bibfield  {author} {\bibinfo {author} {\bibfnamefont {D.}~\bibnamefont
  {Nigg}}, \bibinfo {author} {\bibfnamefont {M.}~\bibnamefont {M{\"u}ller}},
  \bibinfo {author} {\bibfnamefont {E.}~\bibnamefont {Martinez}}, \bibinfo
  {author} {\bibfnamefont {P.}~\bibnamefont {Schindler}}, \bibinfo {author}
  {\bibfnamefont {M.}~\bibnamefont {Hennrich}}, \bibinfo {author}
  {\bibfnamefont {T.}~\bibnamefont {Monz}}, \bibinfo {author} {\bibfnamefont
  {M.}~\bibnamefont {Martin-Delgado}}, \ and\ \bibinfo {author} {\bibfnamefont
  {R.}~\bibnamefont {Blatt}},\ }\href@noop {} {\bibfield  {journal} {\bibinfo
  {journal} {Science}\ }\textbf {\bibinfo {volume} {345}},\ \bibinfo {pages}
  {302} (\bibinfo {year} {2014})}\BibitemShut {NoStop}%
\bibitem [{\citenamefont {Yao}\ \emph {et~al.}(2012)\citenamefont {Yao},
  \citenamefont {Wang}, \citenamefont {Chen}, \citenamefont {Gao},
  \citenamefont {Fowler}, \citenamefont {Raussendorf}, \citenamefont {Chen},
  \citenamefont {Liu}, \citenamefont {Lu}, \citenamefont {Deng}, \citenamefont
  {Chen},\ and\ \citenamefont {Pan}}]{Yao2012}%
  \BibitemOpen
  \bibfield  {author} {\bibinfo {author} {\bibfnamefont {X.-C.}\ \bibnamefont
  {Yao}}, \bibinfo {author} {\bibfnamefont {T.-X.}\ \bibnamefont {Wang}},
  \bibinfo {author} {\bibfnamefont {H.-Z.}\ \bibnamefont {Chen}}, \bibinfo
  {author} {\bibfnamefont {W.-B.}\ \bibnamefont {Gao}}, \bibinfo {author}
  {\bibfnamefont {A.~G.}\ \bibnamefont {Fowler}}, \bibinfo {author}
  {\bibfnamefont {R.}~\bibnamefont {Raussendorf}}, \bibinfo {author}
  {\bibfnamefont {Z.-B.}\ \bibnamefont {Chen}}, \bibinfo {author}
  {\bibfnamefont {N.-L.}\ \bibnamefont {Liu}}, \bibinfo {author} {\bibfnamefont
  {C.-Y.}\ \bibnamefont {Lu}}, \bibinfo {author} {\bibfnamefont {Y.-J.}\
  \bibnamefont {Deng}}, \bibinfo {author} {\bibfnamefont {Y.-A.}\ \bibnamefont
  {Chen}}, \ and\ \bibinfo {author} {\bibfnamefont {J.-W.}\ \bibnamefont
  {Pan}},\ }\href@noop {} {\bibfield  {journal} {\bibinfo  {journal} {Nature}\
  }\textbf {\bibinfo {volume} {482}},\ \bibinfo {pages} {489} (\bibinfo {year}
  {2012})}\BibitemShut {NoStop}%
\bibitem [{\citenamefont {Waldherr}\ \emph {et~al.}(2014)\citenamefont
  {Waldherr}, \citenamefont {Wang}, \citenamefont {Zaiser}, \citenamefont
  {Jamali}, \citenamefont {Schulte-Herbr{\"u}ggen}, \citenamefont {Abe},
  \citenamefont {Ohshima}, \citenamefont {Isoya}, \citenamefont {Du},
  \citenamefont {Neumann} \emph {et~al.}}]{Waldherr2014}%
  \BibitemOpen
  \bibfield  {author} {\bibinfo {author} {\bibfnamefont {G.}~\bibnamefont
  {Waldherr}}, \bibinfo {author} {\bibfnamefont {Y.}~\bibnamefont {Wang}},
  \bibinfo {author} {\bibfnamefont {S.}~\bibnamefont {Zaiser}}, \bibinfo
  {author} {\bibfnamefont {M.}~\bibnamefont {Jamali}}, \bibinfo {author}
  {\bibfnamefont {T.}~\bibnamefont {Schulte-Herbr{\"u}ggen}}, \bibinfo {author}
  {\bibfnamefont {H.}~\bibnamefont {Abe}}, \bibinfo {author} {\bibfnamefont
  {T.}~\bibnamefont {Ohshima}}, \bibinfo {author} {\bibfnamefont
  {J.}~\bibnamefont {Isoya}}, \bibinfo {author} {\bibfnamefont
  {J.}~\bibnamefont {Du}}, \bibinfo {author} {\bibfnamefont {P.}~\bibnamefont
  {Neumann}},  \emph {et~al.},\ }\href@noop {} {\bibfield  {journal} {\bibinfo
  {journal} {Nature}\ }\textbf {\bibinfo {volume} {506}},\ \bibinfo {pages}
  {204} (\bibinfo {year} {2014})}\BibitemShut {NoStop}%
\bibitem [{\citenamefont {Rist\`e}\ \emph
  {et~al.}(2012{\natexlab{a}})\citenamefont {Rist\`e}, \citenamefont {Bultink},
  \citenamefont {Lehnert},\ and\ \citenamefont {DiCarlo}}]{Riste2012b}%
  \BibitemOpen
  \bibfield  {author} {\bibinfo {author} {\bibfnamefont {D.}~\bibnamefont
  {Rist\`e}}, \bibinfo {author} {\bibfnamefont {C.~C.}\ \bibnamefont
  {Bultink}}, \bibinfo {author} {\bibfnamefont {K.~W.}\ \bibnamefont
  {Lehnert}}, \ and\ \bibinfo {author} {\bibfnamefont {L.}~\bibnamefont
  {DiCarlo}},\ }\href {\doibase 10.1103/PhysRevLett.109.240502} {\bibfield
  {journal} {\bibinfo  {journal} {Phys. Rev. Lett.}\ }\textbf {\bibinfo
  {volume} {109}},\ \bibinfo {pages} {240502} (\bibinfo {year}
  {2012}{\natexlab{a}})}\BibitemShut {NoStop}%
\bibitem [{\citenamefont {Vijay}\ \emph {et~al.}(2012)\citenamefont {Vijay},
  \citenamefont {Macklin}, \citenamefont {Slichter}, \citenamefont {Weber},
  \citenamefont {Murch}, \citenamefont {Naik}, \citenamefont {Korotkov},\ and\
  \citenamefont {Siddiqi}}]{Vijay2012}%
  \BibitemOpen
  \bibfield  {author} {\bibinfo {author} {\bibfnamefont {R.}~\bibnamefont
  {Vijay}}, \bibinfo {author} {\bibfnamefont {C.}~\bibnamefont {Macklin}},
  \bibinfo {author} {\bibfnamefont {D.}~\bibnamefont {Slichter}}, \bibinfo
  {author} {\bibfnamefont {S.}~\bibnamefont {Weber}}, \bibinfo {author}
  {\bibfnamefont {K.}~\bibnamefont {Murch}}, \bibinfo {author} {\bibfnamefont
  {R.}~\bibnamefont {Naik}}, \bibinfo {author} {\bibfnamefont {A.~N.}\
  \bibnamefont {Korotkov}}, \ and\ \bibinfo {author} {\bibfnamefont
  {I.}~\bibnamefont {Siddiqi}},\ }\href {http://dx.doi.org/10.1038/nature11505}
  {\bibfield  {journal} {\bibinfo  {journal} {Nature}\ }\textbf {\bibinfo
  {volume} {490}},\ \bibinfo {pages} {77} (\bibinfo {year} {2012})}\BibitemShut
  {NoStop}%
\bibitem [{\citenamefont {Campagne-Ibarcq}\ \emph {et~al.}(2013)\citenamefont
  {Campagne-Ibarcq}, \citenamefont {Flurin}, \citenamefont {Roch},
  \citenamefont {Darson}, \citenamefont {Morfin}, \citenamefont {Mirrahimi},
  \citenamefont {Devoret}, \citenamefont {Mallet},\ and\ \citenamefont
  {Huard}}]{Campagne-Ibarcq2013}%
  \BibitemOpen
  \bibfield  {author} {\bibinfo {author} {\bibfnamefont {P.}~\bibnamefont
  {Campagne-Ibarcq}}, \bibinfo {author} {\bibfnamefont {E.}~\bibnamefont
  {Flurin}}, \bibinfo {author} {\bibfnamefont {N.}~\bibnamefont {Roch}},
  \bibinfo {author} {\bibfnamefont {D.}~\bibnamefont {Darson}}, \bibinfo
  {author} {\bibfnamefont {P.}~\bibnamefont {Morfin}}, \bibinfo {author}
  {\bibfnamefont {M.}~\bibnamefont {Mirrahimi}}, \bibinfo {author}
  {\bibfnamefont {M.~H.}\ \bibnamefont {Devoret}}, \bibinfo {author}
  {\bibfnamefont {F.}~\bibnamefont {Mallet}}, \ and\ \bibinfo {author}
  {\bibfnamefont {B.}~\bibnamefont {Huard}},\ }\href {\doibase
  10.1103/PhysRevX.3.021008} {\bibfield  {journal} {\bibinfo  {journal} {Phys.
  Rev. X}\ }\textbf {\bibinfo {volume} {3}},\ \bibinfo {pages} {021008}
  (\bibinfo {year} {2013})}\BibitemShut {NoStop}%
\bibitem [{\citenamefont {Riste}\ \emph {et~al.}(2013)\citenamefont {Riste},
  \citenamefont {Dukalski}, \citenamefont {Watson}, \citenamefont {de~Lange},
  \citenamefont {Tiggelman}, \citenamefont {Blanter}, \citenamefont {Lehnert},
  \citenamefont {Schouten},\ and\ \citenamefont {DiCarlo}}]{Riste2013Nature}%
  \BibitemOpen
  \bibfield  {author} {\bibinfo {author} {\bibfnamefont {D.}~\bibnamefont
  {Riste}}, \bibinfo {author} {\bibfnamefont {M.}~\bibnamefont {Dukalski}},
  \bibinfo {author} {\bibfnamefont {C.~A.}\ \bibnamefont {Watson}}, \bibinfo
  {author} {\bibfnamefont {G.}~\bibnamefont {de~Lange}}, \bibinfo {author}
  {\bibfnamefont {M.~J.}\ \bibnamefont {Tiggelman}}, \bibinfo {author}
  {\bibfnamefont {Y.~M.}\ \bibnamefont {Blanter}}, \bibinfo {author}
  {\bibfnamefont {K.~W.}\ \bibnamefont {Lehnert}}, \bibinfo {author}
  {\bibfnamefont {R.~N.}\ \bibnamefont {Schouten}}, \ and\ \bibinfo {author}
  {\bibfnamefont {L.}~\bibnamefont {DiCarlo}},\ }\href@noop {} {\bibfield
  {journal} {\bibinfo  {journal} {Nature}\ }\textbf {\bibinfo {volume} {502}},\
  \bibinfo {pages} {350} (\bibinfo {year} {2013})}\BibitemShut {NoStop}%
\bibitem [{\citenamefont {Steffen}\ \emph {et~al.}(2013)\citenamefont
  {Steffen}, \citenamefont {Salathe}, \citenamefont {Oppliger}, \citenamefont
  {Kurpiers}, \citenamefont {Baur}, \citenamefont {Lang}, \citenamefont
  {Eichler}, \citenamefont {Puebla-Hellmann}, \citenamefont {Fedorov},\ and\
  \citenamefont {Wallraff}}]{Steffen2013Nat}%
  \BibitemOpen
  \bibfield  {author} {\bibinfo {author} {\bibfnamefont {L.}~\bibnamefont
  {Steffen}}, \bibinfo {author} {\bibfnamefont {Y.}~\bibnamefont {Salathe}},
  \bibinfo {author} {\bibfnamefont {M.}~\bibnamefont {Oppliger}}, \bibinfo
  {author} {\bibfnamefont {P.}~\bibnamefont {Kurpiers}}, \bibinfo {author}
  {\bibfnamefont {M.}~\bibnamefont {Baur}}, \bibinfo {author} {\bibfnamefont
  {C.}~\bibnamefont {Lang}}, \bibinfo {author} {\bibfnamefont {C.}~\bibnamefont
  {Eichler}}, \bibinfo {author} {\bibfnamefont {G.}~\bibnamefont
  {Puebla-Hellmann}}, \bibinfo {author} {\bibfnamefont {A.}~\bibnamefont
  {Fedorov}}, \ and\ \bibinfo {author} {\bibfnamefont {A.}~\bibnamefont
  {Wallraff}},\ }\href {http://dx.doi.org/10.1038/nature12422} {\bibfield
  {journal} {\bibinfo  {journal} {Nature}\ }\textbf {\bibinfo {volume} {500}},\
  \bibinfo {pages} {319} (\bibinfo {year} {2013})}\BibitemShut {NoStop}%
\bibitem [{\citenamefont {Barends}\ \emph {et~al.}(2014)\citenamefont
  {Barends}, \citenamefont {Kelly}, \citenamefont {Megrant}, \citenamefont
  {Veitia}, \citenamefont {Sank}, \citenamefont {Jeffrey}, \citenamefont
  {White}, \citenamefont {Mutus}, \citenamefont {Fowler}, \citenamefont
  {Campbell}, \citenamefont {Chen}, \citenamefont {Chen}, \citenamefont
  {Chiaro}, \citenamefont {Dunsworth}, \citenamefont {Neill}, \citenamefont
  {O\'Malley}, \citenamefont {Roushan}, \citenamefont {Vainsencher},
  \citenamefont {Wenner}, \citenamefont {Korotkov}, \citenamefont {Cleland},\
  and\ \citenamefont {Martinis}}]{Barends2014Nature}%
  \BibitemOpen
  \bibfield  {author} {\bibinfo {author} {\bibfnamefont {R.}~\bibnamefont
  {Barends}}, \bibinfo {author} {\bibfnamefont {J.}~\bibnamefont {Kelly}},
  \bibinfo {author} {\bibfnamefont {A.}~\bibnamefont {Megrant}}, \bibinfo
  {author} {\bibfnamefont {A.}~\bibnamefont {Veitia}}, \bibinfo {author}
  {\bibfnamefont {D.}~\bibnamefont {Sank}}, \bibinfo {author} {\bibfnamefont
  {E.}~\bibnamefont {Jeffrey}}, \bibinfo {author} {\bibfnamefont {T.~C.}\
  \bibnamefont {White}}, \bibinfo {author} {\bibfnamefont {J.}~\bibnamefont
  {Mutus}}, \bibinfo {author} {\bibfnamefont {A.~G.}\ \bibnamefont {Fowler}},
  \bibinfo {author} {\bibfnamefont {B.}~\bibnamefont {Campbell}}, \bibinfo
  {author} {\bibfnamefont {Y.}~\bibnamefont {Chen}}, \bibinfo {author}
  {\bibfnamefont {Z.}~\bibnamefont {Chen}}, \bibinfo {author} {\bibfnamefont
  {B.}~\bibnamefont {Chiaro}}, \bibinfo {author} {\bibfnamefont
  {A.}~\bibnamefont {Dunsworth}}, \bibinfo {author} {\bibfnamefont
  {C.}~\bibnamefont {Neill}}, \bibinfo {author} {\bibfnamefont
  {P.}~\bibnamefont {O\'Malley}}, \bibinfo {author} {\bibfnamefont
  {P.}~\bibnamefont {Roushan}}, \bibinfo {author} {\bibfnamefont
  {A.}~\bibnamefont {Vainsencher}}, \bibinfo {author} {\bibfnamefont
  {J.}~\bibnamefont {Wenner}}, \bibinfo {author} {\bibfnamefont {A.~N.}\
  \bibnamefont {Korotkov}}, \bibinfo {author} {\bibfnamefont {A.~N.}\
  \bibnamefont {Cleland}}, \ and\ \bibinfo {author} {\bibfnamefont {J.~M.}\
  \bibnamefont {Martinis}},\ }\href {http://dx.doi.org/10.1038/nature13171}
  {\bibfield  {journal} {\bibinfo  {journal} {Nature}\ }\textbf {\bibinfo
  {volume} {508}},\ \bibinfo {pages} {500} (\bibinfo {year}
  {2014})}\BibitemShut {NoStop}%
\bibitem [{\citenamefont {Chow}\ \emph {et~al.}(2014)\citenamefont {Chow},
  \citenamefont {Gambetta}, \citenamefont {Magesan}, \citenamefont {Abraham},
  \citenamefont {Cross}, \citenamefont {Johnson}, \citenamefont {Masluk},
  \citenamefont {Ryan}, \citenamefont {Smolin}, \citenamefont {Srinivasan},\
  and\ \citenamefont {Steffen}}]{Chow2014Nature}%
  \BibitemOpen
  \bibfield  {author} {\bibinfo {author} {\bibfnamefont {J.~M.}\ \bibnamefont
  {Chow}}, \bibinfo {author} {\bibfnamefont {J.~M.}\ \bibnamefont {Gambetta}},
  \bibinfo {author} {\bibfnamefont {E.}~\bibnamefont {Magesan}}, \bibinfo
  {author} {\bibfnamefont {D.~W.}\ \bibnamefont {Abraham}}, \bibinfo {author}
  {\bibfnamefont {A.~W.}\ \bibnamefont {Cross}}, \bibinfo {author}
  {\bibfnamefont {B.~R.}\ \bibnamefont {Johnson}}, \bibinfo {author}
  {\bibfnamefont {N.~A.}\ \bibnamefont {Masluk}}, \bibinfo {author}
  {\bibfnamefont {C.~A.}\ \bibnamefont {Ryan}}, \bibinfo {author}
  {\bibfnamefont {J.~A.}\ \bibnamefont {Smolin}}, \bibinfo {author}
  {\bibfnamefont {S.~J.}\ \bibnamefont {Srinivasan}}, \ and\ \bibinfo {author}
  {\bibfnamefont {M.}~\bibnamefont {Steffen}},\ }\href
  {http://dx.doi.org/10.1038/ncomms5015} {\bibfield  {journal} {\bibinfo
  {journal} {Nat. Commun.}\ }\textbf {\bibinfo {volume} {5}} (\bibinfo {year}
  {2014})}\BibitemShut {NoStop}%
\bibitem [{\citenamefont {Rist{\`e}}\ \emph {et~al.}(2014)\citenamefont
  {Rist{\`e}}, \citenamefont {Poletto}, \citenamefont {Huang}, \citenamefont
  {Bruno}, \citenamefont {Vesterinen}, \citenamefont {Saira},\ and\
  \citenamefont {DiCarlo}}]{Riste2014NatComm}%
  \BibitemOpen
  \bibfield  {author} {\bibinfo {author} {\bibfnamefont {D.}~\bibnamefont
  {Rist{\`e}}}, \bibinfo {author} {\bibfnamefont {S.}~\bibnamefont {Poletto}},
  \bibinfo {author} {\bibfnamefont {M.-Z.}\ \bibnamefont {Huang}}, \bibinfo
  {author} {\bibfnamefont {A.}~\bibnamefont {Bruno}}, \bibinfo {author}
  {\bibfnamefont {V.}~\bibnamefont {Vesterinen}}, \bibinfo {author}
  {\bibfnamefont {O.-P.}\ \bibnamefont {Saira}}, \ and\ \bibinfo {author}
  {\bibfnamefont {L.}~\bibnamefont {DiCarlo}},\ }\href
  {http://arxiv.org/abs/1411.5542} {\bibfield  {journal} {\bibinfo  {journal}
  {Nat. Commun.}\ }\textbf {\bibinfo {volume} {6}} (\bibinfo {year}
  {2014})}\BibitemShut {NoStop}%
\bibitem [{\citenamefont {Kelly}\ \emph {et~al.}(2015)\citenamefont {Kelly},
  \citenamefont {Barends}, \citenamefont {Fowler}, \citenamefont {Megrant},
  \citenamefont {Jeffrey}, \citenamefont {White}, \citenamefont {Sank},
  \citenamefont {Mutus}, \citenamefont {Campbell}, \citenamefont {Chen} \emph
  {et~al.}}]{Kelly2015Nature}%
  \BibitemOpen
  \bibfield  {author} {\bibinfo {author} {\bibfnamefont {J.}~\bibnamefont
  {Kelly}}, \bibinfo {author} {\bibfnamefont {R.}~\bibnamefont {Barends}},
  \bibinfo {author} {\bibfnamefont {A.}~\bibnamefont {Fowler}}, \bibinfo
  {author} {\bibfnamefont {A.}~\bibnamefont {Megrant}}, \bibinfo {author}
  {\bibfnamefont {E.}~\bibnamefont {Jeffrey}}, \bibinfo {author} {\bibfnamefont
  {T.}~\bibnamefont {White}}, \bibinfo {author} {\bibfnamefont
  {D.}~\bibnamefont {Sank}}, \bibinfo {author} {\bibfnamefont {J.}~\bibnamefont
  {Mutus}}, \bibinfo {author} {\bibfnamefont {B.}~\bibnamefont {Campbell}},
  \bibinfo {author} {\bibfnamefont {Y.}~\bibnamefont {Chen}},  \emph {et~al.},\
  }\href@noop {} {\bibfield  {journal} {\bibinfo  {journal} {Nature}\ }\textbf
  {\bibinfo {volume} {519}},\ \bibinfo {pages} {66} (\bibinfo {year}
  {2015})}\BibitemShut {NoStop}%
\bibitem [{\citenamefont {C{\'o}rcoles}\ \emph {et~al.}(2015)\citenamefont
  {C{\'o}rcoles}, \citenamefont {Magesan}, \citenamefont {Srinivasan},
  \citenamefont {Cross}, \citenamefont {Steffen}, \citenamefont {Gambetta},\
  and\ \citenamefont {Chow}}]{Corcoles2015}%
  \BibitemOpen
  \bibfield  {author} {\bibinfo {author} {\bibfnamefont {A.}~\bibnamefont
  {C{\'o}rcoles}}, \bibinfo {author} {\bibfnamefont {E.}~\bibnamefont
  {Magesan}}, \bibinfo {author} {\bibfnamefont {S.~J.}\ \bibnamefont
  {Srinivasan}}, \bibinfo {author} {\bibfnamefont {A.~W.}\ \bibnamefont
  {Cross}}, \bibinfo {author} {\bibfnamefont {M.}~\bibnamefont {Steffen}},
  \bibinfo {author} {\bibfnamefont {J.~M.}\ \bibnamefont {Gambetta}}, \ and\
  \bibinfo {author} {\bibfnamefont {J.~M.}\ \bibnamefont {Chow}},\ }\href@noop
  {} {\bibfield  {journal} {\bibinfo  {journal} {Nat. Commun.}\ }\textbf
  {\bibinfo {volume} {6}} (\bibinfo {year} {2015})}\BibitemShut {NoStop}%
\bibitem [{\citenamefont {Poyatos}\ \emph {et~al.}(1996)\citenamefont
  {Poyatos}, \citenamefont {Cirac},\ and\ \citenamefont
  {Zoller}}]{Poyatos1996}%
  \BibitemOpen
  \bibfield  {author} {\bibinfo {author} {\bibfnamefont {J.~F.}\ \bibnamefont
  {Poyatos}}, \bibinfo {author} {\bibfnamefont {J.~I.}\ \bibnamefont {Cirac}},
  \ and\ \bibinfo {author} {\bibfnamefont {P.}~\bibnamefont {Zoller}},\ }\href
  {\doibase 10.1103/PhysRevLett.77.4728} {\bibfield  {journal} {\bibinfo
  {journal} {Phys. Rev. Lett.}\ }\textbf {\bibinfo {volume} {77}},\ \bibinfo
  {pages} {4728} (\bibinfo {year} {1996})}\BibitemShut {NoStop}%
\bibitem [{\citenamefont {Krauter}\ \emph {et~al.}(2011)\citenamefont
  {Krauter}, \citenamefont {Muschik}, \citenamefont {Jensen}, \citenamefont
  {Wasilewski}, \citenamefont {Petersen}, \citenamefont {Cirac},\ and\
  \citenamefont {Polzik}}]{Krauter2011}%
  \BibitemOpen
  \bibfield  {author} {\bibinfo {author} {\bibfnamefont {H.}~\bibnamefont
  {Krauter}}, \bibinfo {author} {\bibfnamefont {C.~A.}\ \bibnamefont
  {Muschik}}, \bibinfo {author} {\bibfnamefont {K.}~\bibnamefont {Jensen}},
  \bibinfo {author} {\bibfnamefont {W.}~\bibnamefont {Wasilewski}}, \bibinfo
  {author} {\bibfnamefont {J.~M.}\ \bibnamefont {Petersen}}, \bibinfo {author}
  {\bibfnamefont {J.~I.}\ \bibnamefont {Cirac}}, \ and\ \bibinfo {author}
  {\bibfnamefont {E.~S.}\ \bibnamefont {Polzik}},\ }\href {\doibase
  10.1103/PhysRevLett.107.080503} {\bibfield  {journal} {\bibinfo  {journal}
  {Phys. Rev. Lett.}\ }\textbf {\bibinfo {volume} {107}},\ \bibinfo {pages}
  {080503} (\bibinfo {year} {2011})}\BibitemShut {NoStop}%
\bibitem [{\citenamefont {Lin}\ \emph {et~al.}(2013)\citenamefont {Lin},
  \citenamefont {Gaebler}, \citenamefont {Reiter}, \citenamefont {Tan},
  \citenamefont {Bowler}, \citenamefont {S{\o}rensen}, \citenamefont
  {Leibfried},\ and\ \citenamefont {Wineland}}]{lin2013dissipative}%
  \BibitemOpen
  \bibfield  {author} {\bibinfo {author} {\bibfnamefont {Y.}~\bibnamefont
  {Lin}}, \bibinfo {author} {\bibfnamefont {J.}~\bibnamefont {Gaebler}},
  \bibinfo {author} {\bibfnamefont {F.}~\bibnamefont {Reiter}}, \bibinfo
  {author} {\bibfnamefont {T.~R.}\ \bibnamefont {Tan}}, \bibinfo {author}
  {\bibfnamefont {R.}~\bibnamefont {Bowler}}, \bibinfo {author} {\bibfnamefont
  {A.}~\bibnamefont {S{\o}rensen}}, \bibinfo {author} {\bibfnamefont
  {D.}~\bibnamefont {Leibfried}}, \ and\ \bibinfo {author} {\bibfnamefont
  {D.}~\bibnamefont {Wineland}},\ }\href@noop {} {\bibfield  {journal}
  {\bibinfo  {journal} {Nature}\ }\textbf {\bibinfo {volume} {504}},\ \bibinfo
  {pages} {415} (\bibinfo {year} {2013})}\BibitemShut {NoStop}%
\bibitem [{\citenamefont {Kerckhoff}\ \emph {et~al.}(2013)\citenamefont
  {Kerckhoff}, \citenamefont {Andrews}, \citenamefont {Ku}, \citenamefont
  {Kindel}, \citenamefont {Cicak}, \citenamefont {Simmonds},\ and\
  \citenamefont {Lehnert}}]{Kerckhoff2013}%
  \BibitemOpen
  \bibfield  {author} {\bibinfo {author} {\bibfnamefont {J.}~\bibnamefont
  {Kerckhoff}}, \bibinfo {author} {\bibfnamefont {R.~W.}\ \bibnamefont
  {Andrews}}, \bibinfo {author} {\bibfnamefont {H.~S.}\ \bibnamefont {Ku}},
  \bibinfo {author} {\bibfnamefont {W.~F.}\ \bibnamefont {Kindel}}, \bibinfo
  {author} {\bibfnamefont {K.}~\bibnamefont {Cicak}}, \bibinfo {author}
  {\bibfnamefont {R.~W.}\ \bibnamefont {Simmonds}}, \ and\ \bibinfo {author}
  {\bibfnamefont {K.~W.}\ \bibnamefont {Lehnert}},\ }\href {\doibase
  10.1103/PhysRevX.3.021013} {\bibfield  {journal} {\bibinfo  {journal} {Phys.
  Rev. X}\ }\textbf {\bibinfo {volume} {3}},\ \bibinfo {pages} {021013}
  (\bibinfo {year} {2013})}\BibitemShut {NoStop}%
\bibitem [{\citenamefont {Kienzler}\ \emph {et~al.}(2015)\citenamefont
  {Kienzler}, \citenamefont {Lo}, \citenamefont {Keitch}, \citenamefont
  {de~Clercq}, \citenamefont {Leupold}, \citenamefont {Lindenfelser},
  \citenamefont {Marinelli}, \citenamefont {Negnevitsky},\ and\ \citenamefont
  {Home}}]{kienzler2015}%
  \BibitemOpen
  \bibfield  {author} {\bibinfo {author} {\bibfnamefont {D.}~\bibnamefont
  {Kienzler}}, \bibinfo {author} {\bibfnamefont {H.-Y.}\ \bibnamefont {Lo}},
  \bibinfo {author} {\bibfnamefont {B.}~\bibnamefont {Keitch}}, \bibinfo
  {author} {\bibfnamefont {L.}~\bibnamefont {de~Clercq}}, \bibinfo {author}
  {\bibfnamefont {F.}~\bibnamefont {Leupold}}, \bibinfo {author} {\bibfnamefont
  {F.}~\bibnamefont {Lindenfelser}}, \bibinfo {author} {\bibfnamefont
  {M.}~\bibnamefont {Marinelli}}, \bibinfo {author} {\bibfnamefont
  {V.}~\bibnamefont {Negnevitsky}}, \ and\ \bibinfo {author} {\bibfnamefont
  {J.}~\bibnamefont {Home}},\ }\href@noop {} {\bibfield  {journal} {\bibinfo
  {journal} {Science}\ }\textbf {\bibinfo {volume} {347}},\ \bibinfo {pages}
  {53} (\bibinfo {year} {2015})}\BibitemShut {NoStop}%
\bibitem [{\citenamefont {Murch}\ \emph {et~al.}(2012)\citenamefont {Murch},
  \citenamefont {Vool}, \citenamefont {Zhou}, \citenamefont {Weber},
  \citenamefont {Girvin},\ and\ \citenamefont {Siddiqi}}]{Murch2012}%
  \BibitemOpen
  \bibfield  {author} {\bibinfo {author} {\bibfnamefont {K.~W.}\ \bibnamefont
  {Murch}}, \bibinfo {author} {\bibfnamefont {U.}~\bibnamefont {Vool}},
  \bibinfo {author} {\bibfnamefont {D.}~\bibnamefont {Zhou}}, \bibinfo {author}
  {\bibfnamefont {S.~J.}\ \bibnamefont {Weber}}, \bibinfo {author}
  {\bibfnamefont {S.~M.}\ \bibnamefont {Girvin}}, \ and\ \bibinfo {author}
  {\bibfnamefont {I.}~\bibnamefont {Siddiqi}},\ }\href {\doibase
  10.1103/PhysRevLett.109.183602} {\bibfield  {journal} {\bibinfo  {journal}
  {Phys. Rev. Lett.}\ }\textbf {\bibinfo {volume} {109}},\ \bibinfo {pages}
  {183602} (\bibinfo {year} {2012})}\BibitemShut {NoStop}%
\bibitem [{\citenamefont {Geerlings}\ \emph {et~al.}(2013)\citenamefont
  {Geerlings}, \citenamefont {Leghtas}, \citenamefont {Pop}, \citenamefont
  {Shankar}, \citenamefont {Frunzio}, \citenamefont {Schoelkopf}, \citenamefont
  {Mirrahimi},\ and\ \citenamefont {Devoret}}]{Geerlings2013}%
  \BibitemOpen
  \bibfield  {author} {\bibinfo {author} {\bibfnamefont {K.}~\bibnamefont
  {Geerlings}}, \bibinfo {author} {\bibfnamefont {Z.}~\bibnamefont {Leghtas}},
  \bibinfo {author} {\bibfnamefont {I.~M.}\ \bibnamefont {Pop}}, \bibinfo
  {author} {\bibfnamefont {S.}~\bibnamefont {Shankar}}, \bibinfo {author}
  {\bibfnamefont {L.}~\bibnamefont {Frunzio}}, \bibinfo {author} {\bibfnamefont
  {R.~J.}\ \bibnamefont {Schoelkopf}}, \bibinfo {author} {\bibfnamefont
  {M.}~\bibnamefont {Mirrahimi}}, \ and\ \bibinfo {author} {\bibfnamefont
  {M.~H.}\ \bibnamefont {Devoret}},\ }\href {\doibase
  10.1103/PhysRevLett.110.120501} {\bibfield  {journal} {\bibinfo  {journal}
  {Phys. Rev. Lett.}\ }\textbf {\bibinfo {volume} {110}},\ \bibinfo {pages}
  {120501} (\bibinfo {year} {2013})}\BibitemShut {NoStop}%
\bibitem [{\citenamefont {Leghtas}\ \emph {et~al.}(2013)\citenamefont
  {Leghtas}, \citenamefont {Vool}, \citenamefont {Shankar}, \citenamefont
  {Hatridge}, \citenamefont {Girvin}, \citenamefont {Devoret},\ and\
  \citenamefont {Mirrahimi}}]{Leghtas2013}%
  \BibitemOpen
  \bibfield  {author} {\bibinfo {author} {\bibfnamefont {Z.}~\bibnamefont
  {Leghtas}}, \bibinfo {author} {\bibfnamefont {U.}~\bibnamefont {Vool}},
  \bibinfo {author} {\bibfnamefont {S.}~\bibnamefont {Shankar}}, \bibinfo
  {author} {\bibfnamefont {M.}~\bibnamefont {Hatridge}}, \bibinfo {author}
  {\bibfnamefont {S.~M.}\ \bibnamefont {Girvin}}, \bibinfo {author}
  {\bibfnamefont {M.~H.}\ \bibnamefont {Devoret}}, \ and\ \bibinfo {author}
  {\bibfnamefont {M.}~\bibnamefont {Mirrahimi}},\ }\href {\doibase
  10.1103/PhysRevA.88.023849} {\bibfield  {journal} {\bibinfo  {journal} {Phys.
  Rev. A}\ }\textbf {\bibinfo {volume} {88}},\ \bibinfo {pages} {023849}
  (\bibinfo {year} {2013})}\BibitemShut {NoStop}%
\bibitem [{\citenamefont {Shankar}\ \emph {et~al.}(2013)\citenamefont
  {Shankar}, \citenamefont {Hatridge}, \citenamefont {Leghtas}, \citenamefont
  {Sliwa}, \citenamefont {Narla}, \citenamefont {Vool}, \citenamefont {Girvin},
  \citenamefont {Frunzio}, \citenamefont {Mirrahimi},\ and\ \citenamefont
  {Devoret}}]{Shankar2013}%
  \BibitemOpen
  \bibfield  {author} {\bibinfo {author} {\bibfnamefont {S.}~\bibnamefont
  {Shankar}}, \bibinfo {author} {\bibfnamefont {M.}~\bibnamefont {Hatridge}},
  \bibinfo {author} {\bibfnamefont {Z.}~\bibnamefont {Leghtas}}, \bibinfo
  {author} {\bibfnamefont {K.~M.}\ \bibnamefont {Sliwa}}, \bibinfo {author}
  {\bibfnamefont {A.}~\bibnamefont {Narla}}, \bibinfo {author} {\bibfnamefont
  {U.}~\bibnamefont {Vool}}, \bibinfo {author} {\bibfnamefont {S.~M.}\
  \bibnamefont {Girvin}}, \bibinfo {author} {\bibfnamefont {L.}~\bibnamefont
  {Frunzio}}, \bibinfo {author} {\bibfnamefont {M.}~\bibnamefont {Mirrahimi}},
  \ and\ \bibinfo {author} {\bibfnamefont {M.~H.}\ \bibnamefont {Devoret}},\
  }\href {http://dx.doi.org/10.1038/nature12802} {\bibfield  {journal}
  {\bibinfo  {journal} {Nature}\ }\textbf {\bibinfo {volume} {504}},\ \bibinfo
  {pages} {419} (\bibinfo {year} {2013})}\BibitemShut {NoStop}%
\bibitem [{\citenamefont {Leghtas}\ \emph {et~al.}(2015)\citenamefont
  {Leghtas}, \citenamefont {Touzard}, \citenamefont {Pop}, \citenamefont {Kou},
  \citenamefont {Vlastakis}, \citenamefont {Petrenko}, \citenamefont {Sliwa},
  \citenamefont {Narla}, \citenamefont {Shankar}, \citenamefont {Hatridge},
  \citenamefont {Reagor}, \citenamefont {Frunzio}, \citenamefont {Schoelkopf},
  \citenamefont {Mirrahimi},\ and\ \citenamefont {Devoret}}]{Leghtas2015}%
  \BibitemOpen
  \bibfield  {author} {\bibinfo {author} {\bibfnamefont {Z.}~\bibnamefont
  {Leghtas}}, \bibinfo {author} {\bibfnamefont {S.}~\bibnamefont {Touzard}},
  \bibinfo {author} {\bibfnamefont {I.~M.}\ \bibnamefont {Pop}}, \bibinfo
  {author} {\bibfnamefont {A.}~\bibnamefont {Kou}}, \bibinfo {author}
  {\bibfnamefont {B.}~\bibnamefont {Vlastakis}}, \bibinfo {author}
  {\bibfnamefont {A.}~\bibnamefont {Petrenko}}, \bibinfo {author}
  {\bibfnamefont {K.~M.}\ \bibnamefont {Sliwa}}, \bibinfo {author}
  {\bibfnamefont {A.}~\bibnamefont {Narla}}, \bibinfo {author} {\bibfnamefont
  {S.}~\bibnamefont {Shankar}}, \bibinfo {author} {\bibfnamefont {M.~J.}\
  \bibnamefont {Hatridge}}, \bibinfo {author} {\bibfnamefont {M.}~\bibnamefont
  {Reagor}}, \bibinfo {author} {\bibfnamefont {L.}~\bibnamefont {Frunzio}},
  \bibinfo {author} {\bibfnamefont {R.~J.}\ \bibnamefont {Schoelkopf}},
  \bibinfo {author} {\bibfnamefont {M.}~\bibnamefont {Mirrahimi}}, \ and\
  \bibinfo {author} {\bibfnamefont {M.~H.}\ \bibnamefont {Devoret}},\
  }\href@noop {} {\bibfield  {journal} {\bibinfo  {journal} {Science}\ }\textbf
  {\bibinfo {volume} {347}},\ \bibinfo {pages} {853} (\bibinfo {year}
  {2015})}\BibitemShut {NoStop}%
\bibitem [{\citenamefont {Holland}\ \emph {et~al.}(2015)\citenamefont
  {Holland}, \citenamefont {Vlastakis}, \citenamefont {Heeres}, \citenamefont
  {Reagor}, \citenamefont {Vool}, \citenamefont {Leghtas}, \citenamefont
  {Frunzio}, \citenamefont {Kirchmair}, \citenamefont {Devoret}, \citenamefont
  {Mirrahimi} \emph {et~al.}}]{holland2015}%
  \BibitemOpen
  \bibfield  {author} {\bibinfo {author} {\bibfnamefont {E.}~\bibnamefont
  {Holland}}, \bibinfo {author} {\bibfnamefont {B.}~\bibnamefont {Vlastakis}},
  \bibinfo {author} {\bibfnamefont {R.}~\bibnamefont {Heeres}}, \bibinfo
  {author} {\bibfnamefont {M.}~\bibnamefont {Reagor}}, \bibinfo {author}
  {\bibfnamefont {U.}~\bibnamefont {Vool}}, \bibinfo {author} {\bibfnamefont
  {Z.}~\bibnamefont {Leghtas}}, \bibinfo {author} {\bibfnamefont
  {L.}~\bibnamefont {Frunzio}}, \bibinfo {author} {\bibfnamefont
  {G.}~\bibnamefont {Kirchmair}}, \bibinfo {author} {\bibfnamefont
  {M.}~\bibnamefont {Devoret}}, \bibinfo {author} {\bibfnamefont
  {M.}~\bibnamefont {Mirrahimi}},  \emph {et~al.},\ }\href@noop {} {\bibfield
  {journal} {\bibinfo  {journal} {arXiv:1504.03382}\ } (\bibinfo {year}
  {2015})}\BibitemShut {NoStop}%
\bibitem [{\citenamefont {Schindler}\ \emph {et~al.}(2011)\citenamefont
  {Schindler}, \citenamefont {Barreiro}, \citenamefont {Monz}, \citenamefont
  {Nebendahl}, \citenamefont {Nigg}, \citenamefont {Chwalla}, \citenamefont
  {Hennrich},\ and\ \citenamefont {Blatt}}]{Schindler2011}%
  \BibitemOpen
  \bibfield  {author} {\bibinfo {author} {\bibfnamefont {P.}~\bibnamefont
  {Schindler}}, \bibinfo {author} {\bibfnamefont {J.~T.}\ \bibnamefont
  {Barreiro}}, \bibinfo {author} {\bibfnamefont {T.}~\bibnamefont {Monz}},
  \bibinfo {author} {\bibfnamefont {V.}~\bibnamefont {Nebendahl}}, \bibinfo
  {author} {\bibfnamefont {D.}~\bibnamefont {Nigg}}, \bibinfo {author}
  {\bibfnamefont {M.}~\bibnamefont {Chwalla}}, \bibinfo {author} {\bibfnamefont
  {M.}~\bibnamefont {Hennrich}}, \ and\ \bibinfo {author} {\bibfnamefont
  {R.}~\bibnamefont {Blatt}},\ }\href@noop {} {\bibfield  {journal} {\bibinfo
  {journal} {Science}\ }\textbf {\bibinfo {volume} {332}},\ \bibinfo {pages}
  {1059} (\bibinfo {year} {2011})}\BibitemShut {NoStop}%
\bibitem [{\citenamefont {Reed}\ \emph {et~al.}(2012)\citenamefont {Reed},
  \citenamefont {DiCarlo}, \citenamefont {Nigg}, \citenamefont {Sun},
  \citenamefont {Frunzio}, \citenamefont {Girvin},\ and\ \citenamefont
  {Schoelkopf}}]{Reed2012}%
  \BibitemOpen
  \bibfield  {author} {\bibinfo {author} {\bibfnamefont {M.~D.}\ \bibnamefont
  {Reed}}, \bibinfo {author} {\bibfnamefont {L.}~\bibnamefont {DiCarlo}},
  \bibinfo {author} {\bibfnamefont {S.~E.}\ \bibnamefont {Nigg}}, \bibinfo
  {author} {\bibfnamefont {L.}~\bibnamefont {Sun}}, \bibinfo {author}
  {\bibfnamefont {L.}~\bibnamefont {Frunzio}}, \bibinfo {author} {\bibfnamefont
  {S.~M.}\ \bibnamefont {Girvin}}, \ and\ \bibinfo {author} {\bibfnamefont
  {R.~J.}\ \bibnamefont {Schoelkopf}},\ }\href
  {http://dx.doi.org/10.1038/nature10786} {\bibfield  {journal} {\bibinfo
  {journal} {Nature}\ }\textbf {\bibinfo {volume} {482}},\ \bibinfo {pages}
  {382} (\bibinfo {year} {2012})}\BibitemShut {NoStop}%
\bibitem [{\citenamefont {Wiseman}\ and\ \citenamefont
  {Milburn}(1994)}]{Wiseman1994}%
  \BibitemOpen
  \bibfield  {author} {\bibinfo {author} {\bibfnamefont {H.~M.}\ \bibnamefont
  {Wiseman}}\ and\ \bibinfo {author} {\bibfnamefont {G.~J.}\ \bibnamefont
  {Milburn}},\ }\href@noop {} {\bibfield  {journal} {\bibinfo  {journal}
  {Physical review A}\ }\textbf {\bibinfo {volume} {49}},\ \bibinfo {pages}
  {4110} (\bibinfo {year} {1994})}\BibitemShut {NoStop}%
\bibitem [{\citenamefont {Nurdin}\ \emph {et~al.}(2009)\citenamefont {Nurdin},
  \citenamefont {James},\ and\ \citenamefont {Petersen}}]{Nurdin2009}%
  \BibitemOpen
  \bibfield  {author} {\bibinfo {author} {\bibfnamefont {H.~I.}\ \bibnamefont
  {Nurdin}}, \bibinfo {author} {\bibfnamefont {M.~R.}\ \bibnamefont {James}}, \
  and\ \bibinfo {author} {\bibfnamefont {I.~R.}\ \bibnamefont {Petersen}},\
  }\href {\doibase http://dx.doi.org/10.1016/j.automatica.2009.04.018}
  {\bibfield  {journal} {\bibinfo  {journal} {Automatica}\ }\textbf {\bibinfo
  {volume} {45}},\ \bibinfo {pages} {1837 } (\bibinfo {year}
  {2009})}\BibitemShut {NoStop}%
\bibitem [{\citenamefont {Schreier}\ \emph {et~al.}(2008)\citenamefont
  {Schreier}, \citenamefont {Houck}, \citenamefont {Koch}, \citenamefont
  {Schuster}, \citenamefont {Johnson}, \citenamefont {Chow}, \citenamefont
  {Gambetta}, \citenamefont {Majer}, \citenamefont {Frunzio}, \citenamefont
  {Devoret}, \citenamefont {Girvin},\ and\ \citenamefont
  {Schoelkopf}}]{Schreier2008}%
  \BibitemOpen
  \bibfield  {author} {\bibinfo {author} {\bibfnamefont {J.~A.}\ \bibnamefont
  {Schreier}}, \bibinfo {author} {\bibfnamefont {A.~A.}\ \bibnamefont {Houck}},
  \bibinfo {author} {\bibfnamefont {J.}~\bibnamefont {Koch}}, \bibinfo {author}
  {\bibfnamefont {D.~I.}\ \bibnamefont {Schuster}}, \bibinfo {author}
  {\bibfnamefont {B.~R.}\ \bibnamefont {Johnson}}, \bibinfo {author}
  {\bibfnamefont {J.~M.}\ \bibnamefont {Chow}}, \bibinfo {author}
  {\bibfnamefont {J.~M.}\ \bibnamefont {Gambetta}}, \bibinfo {author}
  {\bibfnamefont {J.}~\bibnamefont {Majer}}, \bibinfo {author} {\bibfnamefont
  {L.}~\bibnamefont {Frunzio}}, \bibinfo {author} {\bibfnamefont {M.~H.}\
  \bibnamefont {Devoret}}, \bibinfo {author} {\bibfnamefont {S.~M.}\
  \bibnamefont {Girvin}}, \ and\ \bibinfo {author} {\bibfnamefont {R.~J.}\
  \bibnamefont {Schoelkopf}},\ }\href {\doibase 10.1103/PhysRevB.77.180502}
  {\bibfield  {journal} {\bibinfo  {journal} {Phys. Rev. B}\ }\textbf {\bibinfo
  {volume} {77}},\ \bibinfo {pages} {180502} (\bibinfo {year}
  {2008})}\BibitemShut {NoStop}%
\bibitem [{\citenamefont {Lalumi\`ere}\ \emph {et~al.}(2010)\citenamefont
  {Lalumi\`ere}, \citenamefont {Gambetta},\ and\ \citenamefont
  {Blais}}]{Lalumiere2010}%
  \BibitemOpen
  \bibfield  {author} {\bibinfo {author} {\bibfnamefont {K.}~\bibnamefont
  {Lalumi\`ere}}, \bibinfo {author} {\bibfnamefont {J.~M.}\ \bibnamefont
  {Gambetta}}, \ and\ \bibinfo {author} {\bibfnamefont {A.}~\bibnamefont
  {Blais}},\ }\href {\doibase 10.1103/PhysRevA.81.040301} {\bibfield  {journal}
  {\bibinfo  {journal} {Phys. Rev. A}\ }\textbf {\bibinfo {volume} {81}},\
  \bibinfo {pages} {040301} (\bibinfo {year} {2010})}\BibitemShut {NoStop}%
\bibitem [{\citenamefont {Tornberg}\ and\ \citenamefont
  {Johansson}(2010)}]{Tornberg2010}%
  \BibitemOpen
  \bibfield  {author} {\bibinfo {author} {\bibfnamefont {L.}~\bibnamefont
  {Tornberg}}\ and\ \bibinfo {author} {\bibfnamefont {G.}~\bibnamefont
  {Johansson}},\ }\href {\doibase 10.1103/PhysRevA.82.012329} {\bibfield
  {journal} {\bibinfo  {journal} {Phys. Rev. A}\ }\textbf {\bibinfo {volume}
  {82}},\ \bibinfo {pages} {012329} (\bibinfo {year} {2010})}\BibitemShut
  {NoStop}%
\bibitem [{\citenamefont {Blais}\ \emph {et~al.}(2004)\citenamefont {Blais},
  \citenamefont {Huang}, \citenamefont {Wallraff}, \citenamefont {Girvin},\
  and\ \citenamefont {Schoelkopf}}]{Blais2004}%
  \BibitemOpen
  \bibfield  {author} {\bibinfo {author} {\bibfnamefont {A.}~\bibnamefont
  {Blais}}, \bibinfo {author} {\bibfnamefont {R.-S.}\ \bibnamefont {Huang}},
  \bibinfo {author} {\bibfnamefont {A.}~\bibnamefont {Wallraff}}, \bibinfo
  {author} {\bibfnamefont {S.~M.}\ \bibnamefont {Girvin}}, \ and\ \bibinfo
  {author} {\bibfnamefont {R.~J.}\ \bibnamefont {Schoelkopf}},\ }\href
  {\doibase 10.1103/PhysRevA.69.062320} {\bibfield  {journal} {\bibinfo
  {journal} {Phys. Rev. A}\ }\textbf {\bibinfo {volume} {69}},\ \bibinfo
  {pages} {062320} (\bibinfo {year} {2004})}\BibitemShut {NoStop}%
\bibitem [{\citenamefont {Yamamoto}(2014)}]{Yamamoto2014coherent}%
  \BibitemOpen
  \bibfield  {author} {\bibinfo {author} {\bibfnamefont {N.}~\bibnamefont
  {Yamamoto}},\ }\href@noop {} {\bibfield  {journal} {\bibinfo  {journal}
  {Physical Review X}\ }\textbf {\bibinfo {volume} {4}},\ \bibinfo {pages}
  {041029} (\bibinfo {year} {2014})}\BibitemShut {NoStop}%
\bibitem [{\citenamefont {Jacobs}\ \emph {et~al.}(2014)\citenamefont {Jacobs},
  \citenamefont {Wang},\ and\ \citenamefont {Wiseman}}]{Jacobs2014coherent}%
  \BibitemOpen
  \bibfield  {author} {\bibinfo {author} {\bibfnamefont {K.}~\bibnamefont
  {Jacobs}}, \bibinfo {author} {\bibfnamefont {X.}~\bibnamefont {Wang}}, \ and\
  \bibinfo {author} {\bibfnamefont {H.~M.}\ \bibnamefont {Wiseman}},\
  }\href@noop {} {\bibfield  {journal} {\bibinfo  {journal} {New Journal of
  Physics}\ }\textbf {\bibinfo {volume} {16}},\ \bibinfo {pages} {073036}
  (\bibinfo {year} {2014})}\BibitemShut {NoStop}%
\bibitem [{\citenamefont {Hamerly}\ and\ \citenamefont
  {Mabuchi}(2012)}]{Hamerly2012advantages}%
  \BibitemOpen
  \bibfield  {author} {\bibinfo {author} {\bibfnamefont {R.}~\bibnamefont
  {Hamerly}}\ and\ \bibinfo {author} {\bibfnamefont {H.}~\bibnamefont
  {Mabuchi}},\ }\href@noop {} {\bibfield  {journal} {\bibinfo  {journal}
  {Physical review letters}\ }\textbf {\bibinfo {volume} {109}},\ \bibinfo
  {pages} {173602} (\bibinfo {year} {2012})}\BibitemShut {NoStop}%
\bibitem [{\citenamefont {Moehring}\ \emph {et~al.}(2007)\citenamefont
  {Moehring}, \citenamefont {Maunz}, \citenamefont {Olmschenk}, \citenamefont
  {Younge}, \citenamefont {Matsukevich}, \citenamefont {Duan},\ and\
  \citenamefont {Monroe}}]{Moehring2007}%
  \BibitemOpen
  \bibfield  {author} {\bibinfo {author} {\bibfnamefont {D.}~\bibnamefont
  {Moehring}}, \bibinfo {author} {\bibfnamefont {P.}~\bibnamefont {Maunz}},
  \bibinfo {author} {\bibfnamefont {S.}~\bibnamefont {Olmschenk}}, \bibinfo
  {author} {\bibfnamefont {K.}~\bibnamefont {Younge}}, \bibinfo {author}
  {\bibfnamefont {D.}~\bibnamefont {Matsukevich}}, \bibinfo {author}
  {\bibfnamefont {L.-M.}\ \bibnamefont {Duan}}, \ and\ \bibinfo {author}
  {\bibfnamefont {C.}~\bibnamefont {Monroe}},\ }\href@noop {} {\bibfield
  {journal} {\bibinfo  {journal} {Nature}\ }\textbf {\bibinfo {volume} {449}},\
  \bibinfo {pages} {68} (\bibinfo {year} {2007})}\BibitemShut {NoStop}%
\bibitem [{\citenamefont {Wagenknecht}\ \emph {et~al.}(2010)\citenamefont
  {Wagenknecht}, \citenamefont {Li}, \citenamefont {Reingruber}, \citenamefont
  {Bao}, \citenamefont {Goebel}, \citenamefont {Chen}, \citenamefont {Zhang},
  \citenamefont {Chen},\ and\ \citenamefont {Pan}}]{Wagenknecht2010}%
  \BibitemOpen
  \bibfield  {author} {\bibinfo {author} {\bibfnamefont {C.}~\bibnamefont
  {Wagenknecht}}, \bibinfo {author} {\bibfnamefont {C.-M.}\ \bibnamefont {Li}},
  \bibinfo {author} {\bibfnamefont {A.}~\bibnamefont {Reingruber}}, \bibinfo
  {author} {\bibfnamefont {X.-H.}\ \bibnamefont {Bao}}, \bibinfo {author}
  {\bibfnamefont {A.}~\bibnamefont {Goebel}}, \bibinfo {author} {\bibfnamefont
  {Y.-A.}\ \bibnamefont {Chen}}, \bibinfo {author} {\bibfnamefont
  {Q.}~\bibnamefont {Zhang}}, \bibinfo {author} {\bibfnamefont
  {K.}~\bibnamefont {Chen}}, \ and\ \bibinfo {author} {\bibfnamefont {J.-W.}\
  \bibnamefont {Pan}},\ }\href@noop {} {\bibfield  {journal} {\bibinfo
  {journal} {Nature Photonics}\ }\textbf {\bibinfo {volume} {4}},\ \bibinfo
  {pages} {549} (\bibinfo {year} {2010})}\BibitemShut {NoStop}%
\bibitem [{\citenamefont {Hofmann}\ \emph {et~al.}(2012)\citenamefont
  {Hofmann}, \citenamefont {Krug}, \citenamefont {Ortegel}, \citenamefont
  {G{\'e}rard}, \citenamefont {Weber}, \citenamefont {Rosenfeld},\ and\
  \citenamefont {Weinfurter}}]{Hofmann2012heralded}%
  \BibitemOpen
  \bibfield  {author} {\bibinfo {author} {\bibfnamefont {J.}~\bibnamefont
  {Hofmann}}, \bibinfo {author} {\bibfnamefont {M.}~\bibnamefont {Krug}},
  \bibinfo {author} {\bibfnamefont {N.}~\bibnamefont {Ortegel}}, \bibinfo
  {author} {\bibfnamefont {L.}~\bibnamefont {G{\'e}rard}}, \bibinfo {author}
  {\bibfnamefont {M.}~\bibnamefont {Weber}}, \bibinfo {author} {\bibfnamefont
  {W.}~\bibnamefont {Rosenfeld}}, \ and\ \bibinfo {author} {\bibfnamefont
  {H.}~\bibnamefont {Weinfurter}},\ }\href@noop {} {\bibfield  {journal}
  {\bibinfo  {journal} {Science}\ }\textbf {\bibinfo {volume} {337}},\ \bibinfo
  {pages} {72} (\bibinfo {year} {2012})}\BibitemShut {NoStop}%
\bibitem [{\citenamefont {Johnson}\ \emph {et~al.}(2012)\citenamefont
  {Johnson}, \citenamefont {Macklin}, \citenamefont {Slichter}, \citenamefont
  {Vijay}, \citenamefont {Weingarten}, \citenamefont {Clarke},\ and\
  \citenamefont {Siddiqi}}]{Johnson2012}%
  \BibitemOpen
  \bibfield  {author} {\bibinfo {author} {\bibfnamefont {J.~E.}\ \bibnamefont
  {Johnson}}, \bibinfo {author} {\bibfnamefont {C.}~\bibnamefont {Macklin}},
  \bibinfo {author} {\bibfnamefont {D.~H.}\ \bibnamefont {Slichter}}, \bibinfo
  {author} {\bibfnamefont {R.}~\bibnamefont {Vijay}}, \bibinfo {author}
  {\bibfnamefont {E.~B.}\ \bibnamefont {Weingarten}}, \bibinfo {author}
  {\bibfnamefont {J.}~\bibnamefont {Clarke}}, \ and\ \bibinfo {author}
  {\bibfnamefont {I.}~\bibnamefont {Siddiqi}},\ }\href {\doibase
  10.1103/PhysRevLett.109.050506} {\bibfield  {journal} {\bibinfo  {journal}
  {Phys. Rev. Lett.}\ }\textbf {\bibinfo {volume} {109}},\ \bibinfo {pages}
  {050506} (\bibinfo {year} {2012})}\BibitemShut {NoStop}%
\bibitem [{\citenamefont {Rist\`e}\ \emph
  {et~al.}(2012{\natexlab{b}})\citenamefont {Rist\`e}, \citenamefont {van
  Leeuwen}, \citenamefont {Ku}, \citenamefont {Lehnert},\ and\ \citenamefont
  {DiCarlo}}]{Riste2012}%
  \BibitemOpen
  \bibfield  {author} {\bibinfo {author} {\bibfnamefont {D.}~\bibnamefont
  {Rist\`e}}, \bibinfo {author} {\bibfnamefont {J.~G.}\ \bibnamefont {van
  Leeuwen}}, \bibinfo {author} {\bibfnamefont {H.-S.}\ \bibnamefont {Ku}},
  \bibinfo {author} {\bibfnamefont {K.~W.}\ \bibnamefont {Lehnert}}, \ and\
  \bibinfo {author} {\bibfnamefont {L.}~\bibnamefont {DiCarlo}},\ }\href
  {\doibase 10.1103/PhysRevLett.109.050507} {\bibfield  {journal} {\bibinfo
  {journal} {Phys. Rev. Lett.}\ }\textbf {\bibinfo {volume} {109}},\ \bibinfo
  {pages} {050507} (\bibinfo {year} {2012}{\natexlab{b}})}\BibitemShut
  {NoStop}%
\bibitem [{\citenamefont {Bernien}\ \emph {et~al.}(2013)\citenamefont
  {Bernien}, \citenamefont {Hensen}, \citenamefont {Pfaff}, \citenamefont
  {Koolstra}, \citenamefont {Blok}, \citenamefont {Robledo}, \citenamefont
  {Taminiau}, \citenamefont {Markham}, \citenamefont {Twitchen}, \citenamefont
  {Childress} \emph {et~al.}}]{Bernien2013heralded}%
  \BibitemOpen
  \bibfield  {author} {\bibinfo {author} {\bibfnamefont {H.}~\bibnamefont
  {Bernien}}, \bibinfo {author} {\bibfnamefont {B.}~\bibnamefont {Hensen}},
  \bibinfo {author} {\bibfnamefont {W.}~\bibnamefont {Pfaff}}, \bibinfo
  {author} {\bibfnamefont {G.}~\bibnamefont {Koolstra}}, \bibinfo {author}
  {\bibfnamefont {M.}~\bibnamefont {Blok}}, \bibinfo {author} {\bibfnamefont
  {L.}~\bibnamefont {Robledo}}, \bibinfo {author} {\bibfnamefont
  {T.}~\bibnamefont {Taminiau}}, \bibinfo {author} {\bibfnamefont
  {M.}~\bibnamefont {Markham}}, \bibinfo {author} {\bibfnamefont
  {D.}~\bibnamefont {Twitchen}}, \bibinfo {author} {\bibfnamefont
  {L.}~\bibnamefont {Childress}},  \emph {et~al.},\ }\href@noop {} {\bibfield
  {journal} {\bibinfo  {journal} {Nature}\ }\textbf {\bibinfo {volume} {497}},\
  \bibinfo {pages} {86} (\bibinfo {year} {2013})}\BibitemShut {NoStop}%
\bibitem [{\citenamefont {Paik}\ \emph {et~al.}(2011)\citenamefont {Paik},
  \citenamefont {Schuster}, \citenamefont {Bishop}, \citenamefont {Kirchmair},
  \citenamefont {Catelani}, \citenamefont {Sears}, \citenamefont {Johnson},
  \citenamefont {Reagor}, \citenamefont {Frunzio}, \citenamefont {Glazman},
  \citenamefont {Girvin}, \citenamefont {Devoret},\ and\ \citenamefont
  {Schoelkopf}}]{Paik2011}%
  \BibitemOpen
  \bibfield  {author} {\bibinfo {author} {\bibfnamefont {H.}~\bibnamefont
  {Paik}}, \bibinfo {author} {\bibfnamefont {D.~I.}\ \bibnamefont {Schuster}},
  \bibinfo {author} {\bibfnamefont {L.~S.}\ \bibnamefont {Bishop}}, \bibinfo
  {author} {\bibfnamefont {G.}~\bibnamefont {Kirchmair}}, \bibinfo {author}
  {\bibfnamefont {G.}~\bibnamefont {Catelani}}, \bibinfo {author}
  {\bibfnamefont {A.~P.}\ \bibnamefont {Sears}}, \bibinfo {author}
  {\bibfnamefont {B.~R.}\ \bibnamefont {Johnson}}, \bibinfo {author}
  {\bibfnamefont {M.~J.}\ \bibnamefont {Reagor}}, \bibinfo {author}
  {\bibfnamefont {L.}~\bibnamefont {Frunzio}}, \bibinfo {author} {\bibfnamefont
  {L.~I.}\ \bibnamefont {Glazman}}, \bibinfo {author} {\bibfnamefont {S.~M.}\
  \bibnamefont {Girvin}}, \bibinfo {author} {\bibfnamefont {M.~H.}\
  \bibnamefont {Devoret}}, \ and\ \bibinfo {author} {\bibfnamefont {R.~J.}\
  \bibnamefont {Schoelkopf}},\ }\href {\doibase 10.1103/PhysRevLett.107.240501}
  {\bibfield  {journal} {\bibinfo  {journal} {Phys. Rev. Lett.}\ }\textbf
  {\bibinfo {volume} {107}},\ \bibinfo {pages} {240501} (\bibinfo {year}
  {2011})}\BibitemShut {NoStop}%
\bibitem [{\citenamefont {Schuster}\ \emph {et~al.}(2007)\citenamefont
  {Schuster}, \citenamefont {Houck}, \citenamefont {Schreier}, \citenamefont
  {Wallraff}, \citenamefont {Gambetta}, \citenamefont {Blais}, \citenamefont
  {Frunzio}, \citenamefont {Majer}, \citenamefont {Johnson}, \citenamefont
  {Devoret}, \citenamefont {Girvin},\ and\ \citenamefont
  {Schoelkopf}}]{Schuster2007}%
  \BibitemOpen
  \bibfield  {author} {\bibinfo {author} {\bibfnamefont {D.~I.}\ \bibnamefont
  {Schuster}}, \bibinfo {author} {\bibfnamefont {A.~A.}\ \bibnamefont {Houck}},
  \bibinfo {author} {\bibfnamefont {J.~A.}\ \bibnamefont {Schreier}}, \bibinfo
  {author} {\bibfnamefont {A.}~\bibnamefont {Wallraff}}, \bibinfo {author}
  {\bibfnamefont {J.~M.}\ \bibnamefont {Gambetta}}, \bibinfo {author}
  {\bibfnamefont {A.}~\bibnamefont {Blais}}, \bibinfo {author} {\bibfnamefont
  {L.}~\bibnamefont {Frunzio}}, \bibinfo {author} {\bibfnamefont
  {J.}~\bibnamefont {Majer}}, \bibinfo {author} {\bibfnamefont
  {B.}~\bibnamefont {Johnson}}, \bibinfo {author} {\bibfnamefont {M.~H.}\
  \bibnamefont {Devoret}}, \bibinfo {author} {\bibfnamefont {S.~M.}\
  \bibnamefont {Girvin}}, \ and\ \bibinfo {author} {\bibfnamefont {R.~J.}\
  \bibnamefont {Schoelkopf}},\ }\href {http://dx.doi.org/10.1038/nature05461}
  {\bibfield  {journal} {\bibinfo  {journal} {Nature}\ }\textbf {\bibinfo
  {volume} {445}},\ \bibinfo {pages} {515} (\bibinfo {year}
  {2007})}\BibitemShut {NoStop}%
\bibitem [{\citenamefont {Bergeal}\ \emph {et~al.}(2010)\citenamefont
  {Bergeal}, \citenamefont {Schackert}, \citenamefont {Metcalfe}, \citenamefont
  {Vijay}, \citenamefont {Manucharyan}, \citenamefont {Frunzio}, \citenamefont
  {Prober}, \citenamefont {Schoelkopf}, \citenamefont {Girvin},\ and\
  \citenamefont {Devoret}}]{Bergeal2010a}%
  \BibitemOpen
  \bibfield  {author} {\bibinfo {author} {\bibfnamefont {N.}~\bibnamefont
  {Bergeal}}, \bibinfo {author} {\bibfnamefont {F.}~\bibnamefont {Schackert}},
  \bibinfo {author} {\bibfnamefont {M.}~\bibnamefont {Metcalfe}}, \bibinfo
  {author} {\bibfnamefont {R.}~\bibnamefont {Vijay}}, \bibinfo {author}
  {\bibfnamefont {V.~E.}\ \bibnamefont {Manucharyan}}, \bibinfo {author}
  {\bibfnamefont {L.}~\bibnamefont {Frunzio}}, \bibinfo {author} {\bibfnamefont
  {D.~E.}\ \bibnamefont {Prober}}, \bibinfo {author} {\bibfnamefont {R.~J.}\
  \bibnamefont {Schoelkopf}}, \bibinfo {author} {\bibfnamefont {S.~M.}\
  \bibnamefont {Girvin}}, \ and\ \bibinfo {author} {\bibfnamefont {M.~H.}\
  \bibnamefont {Devoret}},\ }\href {http://dx.doi.org/10.1038/nature09035}
  {\bibfield  {journal} {\bibinfo  {journal} {Nature}\ }\textbf {\bibinfo
  {volume} {465}},\ \bibinfo {pages} {64} (\bibinfo {year} {2010})}\BibitemShut
  {NoStop}%
\bibitem [{\citenamefont {Hatridge}\ \emph {et~al.}(2013)\citenamefont
  {Hatridge}, \citenamefont {Shankar}, \citenamefont {Mirrahimi}, \citenamefont
  {Schackert}, \citenamefont {Geerlings}, \citenamefont {Brecht}, \citenamefont
  {Sliwa}, \citenamefont {Abdo}, \citenamefont {Frunzio}, \citenamefont
  {Girvin}, \citenamefont {Schoelkopf},\ and\ \citenamefont
  {Devoret}}]{Hatridge2013}%
  \BibitemOpen
  \bibfield  {author} {\bibinfo {author} {\bibfnamefont {M.}~\bibnamefont
  {Hatridge}}, \bibinfo {author} {\bibfnamefont {S.}~\bibnamefont {Shankar}},
  \bibinfo {author} {\bibfnamefont {M.}~\bibnamefont {Mirrahimi}}, \bibinfo
  {author} {\bibfnamefont {F.}~\bibnamefont {Schackert}}, \bibinfo {author}
  {\bibfnamefont {K.}~\bibnamefont {Geerlings}}, \bibinfo {author}
  {\bibfnamefont {T.}~\bibnamefont {Brecht}}, \bibinfo {author} {\bibfnamefont
  {K.~M.}\ \bibnamefont {Sliwa}}, \bibinfo {author} {\bibfnamefont
  {B.}~\bibnamefont {Abdo}}, \bibinfo {author} {\bibfnamefont {L.}~\bibnamefont
  {Frunzio}}, \bibinfo {author} {\bibfnamefont {S.~M.}\ \bibnamefont {Girvin}},
  \bibinfo {author} {\bibfnamefont {R.~J.}\ \bibnamefont {Schoelkopf}}, \ and\
  \bibinfo {author} {\bibfnamefont {M.~H.}\ \bibnamefont {Devoret}},\ }\href
  {http://dx.doi.org/10.1126/science.1226897} {\bibfield  {journal} {\bibinfo
  {journal} {Science}\ }\textbf {\bibinfo {volume} {339}},\ \bibinfo {pages}
  {178} (\bibinfo {year} {2013})}\BibitemShut {NoStop}%
\bibitem [{Note1()}]{Note1}%
  \BibitemOpen
  \bibinfo {note} {X6-1000M from Innovative Integration}\BibitemShut {NoStop}%
\bibitem [{\citenamefont {Filipp}\ \emph {et~al.}(2009)\citenamefont {Filipp},
  \citenamefont {Maurer}, \citenamefont {Leek}, \citenamefont {Baur},
  \citenamefont {Bianchetti}, \citenamefont {Fink}, \citenamefont {G\"oppl},
  \citenamefont {Steffen}, \citenamefont {Gambetta}, \citenamefont {Blais},\
  and\ \citenamefont {Wallraff}}]{Filipp2009}%
  \BibitemOpen
  \bibfield  {author} {\bibinfo {author} {\bibfnamefont {S.}~\bibnamefont
  {Filipp}}, \bibinfo {author} {\bibfnamefont {P.}~\bibnamefont {Maurer}},
  \bibinfo {author} {\bibfnamefont {P.~J.}\ \bibnamefont {Leek}}, \bibinfo
  {author} {\bibfnamefont {M.}~\bibnamefont {Baur}}, \bibinfo {author}
  {\bibfnamefont {R.}~\bibnamefont {Bianchetti}}, \bibinfo {author}
  {\bibfnamefont {J.~M.}\ \bibnamefont {Fink}}, \bibinfo {author}
  {\bibfnamefont {M.}~\bibnamefont {G\"oppl}}, \bibinfo {author} {\bibfnamefont
  {L.}~\bibnamefont {Steffen}}, \bibinfo {author} {\bibfnamefont {J.~M.}\
  \bibnamefont {Gambetta}}, \bibinfo {author} {\bibfnamefont {A.}~\bibnamefont
  {Blais}}, \ and\ \bibinfo {author} {\bibfnamefont {A.}~\bibnamefont
  {Wallraff}},\ }\href {\doibase 10.1103/PhysRevLett.102.200402} {\bibfield
  {journal} {\bibinfo  {journal} {Phys. Rev. Lett.}\ }\textbf {\bibinfo
  {volume} {102}},\ \bibinfo {pages} {200402} (\bibinfo {year}
  {2009})}\BibitemShut {NoStop}%
\bibitem [{\citenamefont {Chow}\ \emph {et~al.}(2010)\citenamefont {Chow},
  \citenamefont {DiCarlo}, \citenamefont {Gambetta}, \citenamefont
  {Nunnenkamp}, \citenamefont {Bishop}, \citenamefont {Frunzio}, \citenamefont
  {Devoret}, \citenamefont {Girvin},\ and\ \citenamefont
  {Schoelkopf}}]{Chow2010}%
  \BibitemOpen
  \bibfield  {author} {\bibinfo {author} {\bibfnamefont {J.~M.}\ \bibnamefont
  {Chow}}, \bibinfo {author} {\bibfnamefont {L.}~\bibnamefont {DiCarlo}},
  \bibinfo {author} {\bibfnamefont {J.~M.}\ \bibnamefont {Gambetta}}, \bibinfo
  {author} {\bibfnamefont {A.}~\bibnamefont {Nunnenkamp}}, \bibinfo {author}
  {\bibfnamefont {L.~S.}\ \bibnamefont {Bishop}}, \bibinfo {author}
  {\bibfnamefont {L.}~\bibnamefont {Frunzio}}, \bibinfo {author} {\bibfnamefont
  {M.~H.}\ \bibnamefont {Devoret}}, \bibinfo {author} {\bibfnamefont {S.~M.}\
  \bibnamefont {Girvin}}, \ and\ \bibinfo {author} {\bibfnamefont {R.~J.}\
  \bibnamefont {Schoelkopf}},\ }\href {\doibase 10.1103/PhysRevA.81.062325}
  {\bibfield  {journal} {\bibinfo  {journal} {Phys. Rev. A}\ }\textbf {\bibinfo
  {volume} {81}},\ \bibinfo {pages} {062325} (\bibinfo {year}
  {2010})}\BibitemShut {NoStop}%
\bibitem [{\citenamefont {Mirrahimi}\ \emph {et~al.}(2014)\citenamefont
  {Mirrahimi}, \citenamefont {Leghtas}, \citenamefont {Albert}, \citenamefont
  {Touzard}, \citenamefont {Schoelkopf}, \citenamefont {Jiang},\ and\
  \citenamefont {Devoret}}]{Mazyar2014}%
  \BibitemOpen
  \bibfield  {author} {\bibinfo {author} {\bibfnamefont {M.}~\bibnamefont
  {Mirrahimi}}, \bibinfo {author} {\bibfnamefont {Z.}~\bibnamefont {Leghtas}},
  \bibinfo {author} {\bibfnamefont {V.~V.}\ \bibnamefont {Albert}}, \bibinfo
  {author} {\bibfnamefont {S.}~\bibnamefont {Touzard}}, \bibinfo {author}
  {\bibfnamefont {R.~J.}\ \bibnamefont {Schoelkopf}}, \bibinfo {author}
  {\bibfnamefont {L.}~\bibnamefont {Jiang}}, \ and\ \bibinfo {author}
  {\bibfnamefont {M.~H.}\ \bibnamefont {Devoret}},\ }\href
  {http://stacks.iop.org/1367-2630/16/i=4/a=045014} {\bibfield  {journal}
  {\bibinfo  {journal} {New Journal of Physics}\ }\textbf {\bibinfo {volume}
  {16}},\ \bibinfo {pages} {045014} (\bibinfo {year} {2014})}\BibitemShut
  {NoStop}%
\bibitem [{\citenamefont {Sun}\ \emph {et~al.}(2014)\citenamefont {Sun},
  \citenamefont {Petrenko}, \citenamefont {Leghtas}, \citenamefont {Vlastakis},
  \citenamefont {Kirchmair}, \citenamefont {Sliwa}, \citenamefont {Narla},
  \citenamefont {Hatridge}, \citenamefont {Shankar}, \citenamefont {Blumoff},
  \citenamefont {Frunzio}, \citenamefont {Mirrahimi}, \citenamefont {Devoret},\
  and\ \citenamefont {Schoelkopf}}]{Sun2014}%
  \BibitemOpen
  \bibfield  {author} {\bibinfo {author} {\bibfnamefont {L.}~\bibnamefont
  {Sun}}, \bibinfo {author} {\bibfnamefont {A.}~\bibnamefont {Petrenko}},
  \bibinfo {author} {\bibfnamefont {Z.}~\bibnamefont {Leghtas}}, \bibinfo
  {author} {\bibfnamefont {B.}~\bibnamefont {Vlastakis}}, \bibinfo {author}
  {\bibfnamefont {G.}~\bibnamefont {Kirchmair}}, \bibinfo {author}
  {\bibfnamefont {K.~M.}\ \bibnamefont {Sliwa}}, \bibinfo {author}
  {\bibfnamefont {A.}~\bibnamefont {Narla}}, \bibinfo {author} {\bibfnamefont
  {M.}~\bibnamefont {Hatridge}}, \bibinfo {author} {\bibfnamefont
  {S.}~\bibnamefont {Shankar}}, \bibinfo {author} {\bibfnamefont
  {J.}~\bibnamefont {Blumoff}}, \bibinfo {author} {\bibfnamefont
  {L.}~\bibnamefont {Frunzio}}, \bibinfo {author} {\bibfnamefont
  {M.}~\bibnamefont {Mirrahimi}}, \bibinfo {author} {\bibfnamefont {M.~H.}\
  \bibnamefont {Devoret}}, \ and\ \bibinfo {author} {\bibfnamefont {R.~J.}\
  \bibnamefont {Schoelkopf}},\ }\href {http://dx.doi.org/10.1038/nature13436}
  {\bibfield  {journal} {\bibinfo  {journal} {Nature}\ }\textbf {\bibinfo
  {volume} {511}},\ \bibinfo {pages} {444} (\bibinfo {year}
  {2014})}\BibitemShut {NoStop}%
\bibitem [{\citenamefont {Kapit}\ \emph {et~al.}(2015)\citenamefont {Kapit},
  \citenamefont {Chalker},\ and\ \citenamefont {Simon}}]{Kapit2015PRA}%
  \BibitemOpen
  \bibfield  {author} {\bibinfo {author} {\bibfnamefont {E.}~\bibnamefont
  {Kapit}}, \bibinfo {author} {\bibfnamefont {J.~T.}\ \bibnamefont {Chalker}},
  \ and\ \bibinfo {author} {\bibfnamefont {S.~H.}\ \bibnamefont {Simon}},\
  }\href {\doibase 10.1103/PhysRevA.91.062324} {\bibfield  {journal} {\bibinfo
  {journal} {Phys. Rev. A}\ }\textbf {\bibinfo {volume} {91}},\ \bibinfo
  {pages} {062324} (\bibinfo {year} {2015})}\BibitemShut {NoStop}%
\end{thebibliography}

\begin{thebibliography}{7}%
\makeatletter
\providecommand \@ifxundefined [1]{%
 \@ifx{#1\undefined}
}%
\providecommand \@ifnum [1]{%
 \ifnum #1\expandafter \@firstoftwo
 \else \expandafter \@secondoftwo
 \fi
}%
\providecommand \@ifx [1]{%
 \ifx #1\expandafter \@firstoftwo
 \else \expandafter \@secondoftwo
 \fi
}%
\providecommand \natexlab [1]{#1}%
\providecommand \enquote  [1]{``#1''}%
\providecommand \bibnamefont  [1]{#1}%
\providecommand \bibfnamefont [1]{#1}%
\providecommand \citenamefont [1]{#1}%
\providecommand \href@noop [0]{\@secondoftwo}%
\providecommand \href [0]{\begingroup \@sanitize@url \@href}%
\providecommand \@href[1]{\@@startlink{#1}\@@href}%
\providecommand \@@href[1]{\endgroup#1\@@endlink}%
\providecommand \@sanitize@url [0]{\catcode `\\12\catcode `\$12\catcode
  `\&12\catcode `\#12\catcode `\^12\catcode `\_12\catcode `\%12\relax}%
\providecommand \@@startlink[1]{}%
\providecommand \@@endlink[0]{}%
\providecommand \url  [0]{\begingroup\@sanitize@url \@url }%
\providecommand \@url [1]{\endgroup\@href {#1}{\urlprefix }}%
\providecommand \urlprefix  [0]{URL }%
\providecommand \Eprint [0]{\href }%
\providecommand \doibase [0]{http://dx.doi.org/}%
\providecommand \selectlanguage [0]{\@gobble}%
\providecommand \bibinfo  [0]{\@secondoftwo}%
\providecommand \bibfield  [0]{\@secondoftwo}%
\providecommand \translation [1]{[#1]}%
\providecommand \BibitemOpen [0]{}%
\providecommand \bibitemStop [0]{}%
\providecommand \bibitemNoStop [0]{.\EOS\space}%
\providecommand \EOS [0]{\spacefactor3000\relax}%
\providecommand \BibitemShut  [1]{\csname bibitem#1\endcsname}%
\let\auto@bib@innerbib\@empty
\bibitem [{\citenamefont {Riste}\ \emph {et~al.}(2013)\citenamefont {Riste},
  \citenamefont {Dukalski}, \citenamefont {Watson}, \citenamefont {de~Lange},
  \citenamefont {Tiggelman}, \citenamefont {Blanter}, \citenamefont {Lehnert},
  \citenamefont {Schouten},\ and\ \citenamefont {DiCarlo}}]{Riste2013Nature}%
  \BibitemOpen
  \bibfield  {author} {\bibinfo {author} {\bibfnamefont {D.}~\bibnamefont
  {Riste}}, \bibinfo {author} {\bibfnamefont {M.}~\bibnamefont {Dukalski}},
  \bibinfo {author} {\bibfnamefont {C.~A.}\ \bibnamefont {Watson}}, \bibinfo
  {author} {\bibfnamefont {G.}~\bibnamefont {de~Lange}}, \bibinfo {author}
  {\bibfnamefont {M.~J.}\ \bibnamefont {Tiggelman}}, \bibinfo {author}
  {\bibfnamefont {Y.~M.}\ \bibnamefont {Blanter}}, \bibinfo {author}
  {\bibfnamefont {K.~W.}\ \bibnamefont {Lehnert}}, \bibinfo {author}
  {\bibfnamefont {R.~N.}\ \bibnamefont {Schouten}}, \ and\ \bibinfo {author}
  {\bibfnamefont {L.}~\bibnamefont {DiCarlo}},\ }\href@noop {} {\bibfield
  {journal} {\bibinfo  {journal} {Nature}\ }\textbf {\bibinfo {volume} {502}},\
  \bibinfo {pages} {350} (\bibinfo {year} {2013})}\BibitemShut {NoStop}%
\bibitem [{\citenamefont {Tornberg}\ and\ \citenamefont
  {Johansson}(2010)}]{Tornberg2010}%
  \BibitemOpen
  \bibfield  {author} {\bibinfo {author} {\bibfnamefont {L.}~\bibnamefont
  {Tornberg}}\ and\ \bibinfo {author} {\bibfnamefont {G.}~\bibnamefont
  {Johansson}},\ }\href {\doibase 10.1103/PhysRevA.82.012329} {\bibfield
  {journal} {\bibinfo  {journal} {Phys. Rev. A}\ }\textbf {\bibinfo {volume}
  {82}},\ \bibinfo {pages} {012329} (\bibinfo {year} {2010})}\BibitemShut
  {NoStop}%
\bibitem [{\citenamefont {Silveri}(2015)}]{Matti2015_}%
  \BibitemOpen
  \bibfield  {author} {\bibinfo {author} {\bibfnamefont {M.}~\bibnamefont
  {Silveri}},\ }\href@noop {} {\bibfield  {journal} {\bibinfo  {journal} {in
  preparation}\ } (\bibinfo {year} {2015})}\BibitemShut {NoStop}%
\bibitem [{\citenamefont {Gambetta}\ \emph {et~al.}(2007)\citenamefont
  {Gambetta}, \citenamefont {Braff}, \citenamefont {Wallraff}, \citenamefont
  {Girvin},\ and\ \citenamefont {Schoelkopf}}]{Gambetta2007}%
  \BibitemOpen
  \bibfield  {author} {\bibinfo {author} {\bibfnamefont {J.}~\bibnamefont
  {Gambetta}}, \bibinfo {author} {\bibfnamefont {W.~A.}\ \bibnamefont {Braff}},
  \bibinfo {author} {\bibfnamefont {A.}~\bibnamefont {Wallraff}}, \bibinfo
  {author} {\bibfnamefont {S.~M.}\ \bibnamefont {Girvin}}, \ and\ \bibinfo
  {author} {\bibfnamefont {R.~J.}\ \bibnamefont {Schoelkopf}},\ }\href
  {\doibase 10.1103/PhysRevA.76.012325} {\bibfield  {journal} {\bibinfo
  {journal} {Phys. Rev. A}\ }\textbf {\bibinfo {volume} {76}},\ \bibinfo
  {pages} {012325} (\bibinfo {year} {2007})}\BibitemShut {NoStop}%
\bibitem [{\citenamefont {Magesan}\ \emph {et~al.}(2015)\citenamefont
  {Magesan}, \citenamefont {Gambetta}, \citenamefont {C\'orcoles},\ and\
  \citenamefont {Chow}}]{Magesan2015}%
  \BibitemOpen
  \bibfield  {author} {\bibinfo {author} {\bibfnamefont {E.}~\bibnamefont
  {Magesan}}, \bibinfo {author} {\bibfnamefont {J.~M.}\ \bibnamefont
  {Gambetta}}, \bibinfo {author} {\bibfnamefont {A.~D.}\ \bibnamefont
  {C\'orcoles}}, \ and\ \bibinfo {author} {\bibfnamefont {J.~M.}\ \bibnamefont
  {Chow}},\ }\href {\doibase 10.1103/PhysRevLett.114.200501} {\bibfield
  {journal} {\bibinfo  {journal} {Phys. Rev. Lett.}\ }\textbf {\bibinfo
  {volume} {114}},\ \bibinfo {pages} {200501} (\bibinfo {year}
  {2015})}\BibitemShut {NoStop}%
\bibitem [{\citenamefont {Leghtas}\ \emph {et~al.}(2013)\citenamefont
  {Leghtas}, \citenamefont {Vool}, \citenamefont {Shankar}, \citenamefont
  {Hatridge}, \citenamefont {Girvin}, \citenamefont {Devoret},\ and\
  \citenamefont {Mirrahimi}}]{Leghtas2013}%
  \BibitemOpen
  \bibfield  {author} {\bibinfo {author} {\bibfnamefont {Z.}~\bibnamefont
  {Leghtas}}, \bibinfo {author} {\bibfnamefont {U.}~\bibnamefont {Vool}},
  \bibinfo {author} {\bibfnamefont {S.}~\bibnamefont {Shankar}}, \bibinfo
  {author} {\bibfnamefont {M.}~\bibnamefont {Hatridge}}, \bibinfo {author}
  {\bibfnamefont {S.~M.}\ \bibnamefont {Girvin}}, \bibinfo {author}
  {\bibfnamefont {M.~H.}\ \bibnamefont {Devoret}}, \ and\ \bibinfo {author}
  {\bibfnamefont {M.}~\bibnamefont {Mirrahimi}},\ }\href {\doibase
  10.1103/PhysRevA.88.023849} {\bibfield  {journal} {\bibinfo  {journal} {Phys.
  Rev. A}\ }\textbf {\bibinfo {volume} {88}},\ \bibinfo {pages} {023849}
  (\bibinfo {year} {2013})}\BibitemShut {NoStop}%
\bibitem [{\citenamefont {Hatridge}\ \emph {et~al.}(2013)\citenamefont
  {Hatridge}, \citenamefont {Shankar}, \citenamefont {Mirrahimi}, \citenamefont
  {Schackert}, \citenamefont {Geerlings}, \citenamefont {Brecht}, \citenamefont
  {Sliwa}, \citenamefont {Abdo}, \citenamefont {Frunzio}, \citenamefont
  {Girvin}, \citenamefont {Schoelkopf},\ and\ \citenamefont
  {Devoret}}]{Hatridge2013}%
  \BibitemOpen
  \bibfield  {author} {\bibinfo {author} {\bibfnamefont {M.}~\bibnamefont
  {Hatridge}}, \bibinfo {author} {\bibfnamefont {S.}~\bibnamefont {Shankar}},
  \bibinfo {author} {\bibfnamefont {M.}~\bibnamefont {Mirrahimi}}, \bibinfo
  {author} {\bibfnamefont {F.}~\bibnamefont {Schackert}}, \bibinfo {author}
  {\bibfnamefont {K.}~\bibnamefont {Geerlings}}, \bibinfo {author}
  {\bibfnamefont {T.}~\bibnamefont {Brecht}}, \bibinfo {author} {\bibfnamefont
  {K.~M.}\ \bibnamefont {Sliwa}}, \bibinfo {author} {\bibfnamefont
  {B.}~\bibnamefont {Abdo}}, \bibinfo {author} {\bibfnamefont {L.}~\bibnamefont
  {Frunzio}}, \bibinfo {author} {\bibfnamefont {S.~M.}\ \bibnamefont {Girvin}},
  \bibinfo {author} {\bibfnamefont {R.~J.}\ \bibnamefont {Schoelkopf}}, \ and\
  \bibinfo {author} {\bibfnamefont {M.~H.}\ \bibnamefont {Devoret}},\ }\href
  {http://dx.doi.org/10.1126/science.1226897} {\bibfield  {journal} {\bibinfo
  {journal} {Science}\ }\textbf {\bibinfo {volume} {339}},\ \bibinfo {pages}
  {178} (\bibinfo {year} {2013})}\BibitemShut {NoStop}%
\end{thebibliography}

\clearpage
FIG.~S.1. Setup of the experiment. The qubit-cavity system was placed at the base stage of a dilution refrigerator (Oxford Triton200) below 20 mK. On the input side: Two FPGAs (Innovative Integration X6-1000M) generated the pulse envelopes in I and Q quadratures to modulate the qubit drives at $f_{\textrm{Alice}}$ and $f_{\textrm{Bob}}$ frequencies (Vaunix Labbrick LMS-802) for Alice and Bob, respectively. The I/Q modulations were output by the FPGAs with one and two single-sideband modulations in MB and DD (so that both zero-photon and n-photon qubit frequencies were addressed), respectively. The microwave frequency drives and I/Q modulation were mixed by IQ mixers. Two Agilent N5183 microwave generators produced the cavity drives at $f_{ee}$ and $f_{gg}$ respectively. The cavity drives were also pulsed by the FPGAs. On the output side: the transmitted signal through the cavity was directed by two circulators to the JPC for nearly quantum-limited amplification. It was further amplified at the 3 K stage by a cryogenic HEMT amplifier. After additional room temperature amplification, the signal was split into two interferometric setups for readouts at the $f_{gg}$ and $f_{ee}$ frequencies, respectively. In each of the interferometers, the signal arriving from the fridge was mixed with a local oscillator set $+50$ MHz away to produce a down-converted signal at 50 MHz. A copy of the cavity drive that did not go through the fridge was also down-converted in the same manner to produce a reference. Finally, the two signals with their respective references were sent to the analog-to-digital-converters on the FPGA boards for digitization and further demodulated inside the FPGAs. The two FPGAs jointly estimate the qubits' state by communicating their results with each other. Along the input and output lines, attenuators, low pass filters and homemade Eccosorb filters were placed at various stages to protect the qubits from thermal noise and undesired microwave and optical frequency radiation.  Moreover, the cavity and JPC were shielded from stray magnetic fields by aluminum and cryogenic $\mu$-metal (Amumetal A4K) shields.
\clearpage

\clearpage

\end{document}